\documentclass[aps,prb,showpacs,twocolumn,superscriptaddress]{revtex4}
\usepackage{bm,color,amsmath,amssymb,mathrsfs,latexsym,graphicx,psfrag}

\newcommand{\bra}[1]{\left\langle#1\right|}
\newcommand{\ket}[1]{\left|#1\right\rangle}
\newcommand{\Tr}[1]{\mbox{Tr}\left( #1 \right)}

\newcommand{\ev}[1]{\langle #1 \rangle}

\begin{document}
\title{Stochastic Mean-Field Theory: Method and Application to the Disordered Bose-Hubbard Model at Finite Temperature and Speckle Disorder}

\author{Ulf Bissbort}
\affiliation{Institut f\"ur Theoretische Physik, Johann Wolfgang Goethe-Universit\"at, 60438 Frankfurt/Main, Germany}
\author{Ronny Thomale}
\affiliation{Department of Physics, Princeton University, Princeton, NJ 08544}
\author{Walter Hofstetter}
\affiliation{Institut f\"ur Theoretische Physik, Johann Wolfgang Goethe-Universit\"at, 60438 Frankfurt/Main, Germany}

\pagestyle{plain}
\date{\today}

\begin{abstract}
We discuss the stochastic mean-field theory (SMFT) method which is a new
approach for describing disordered Bose systems in the thermodynamic
limit including localization and dimensional effects. We explicate
the method in detail and apply it to the disordered Bose-Hubbard
model at finite temperature, with on-site box disorder, as well as
experimentally relevant unbounded speckle disorder. We find that
disorder-induced condensation and reentrant behavior at constant
filling are only possible at low temperatures, beyond
the reach of current experiments~\cite{Pasienski-arXiv0908.1182}.
Including off-diagonal hopping disorder as well, we
investigate its effect on the phase diagram in addition to
pure on-site disorder.  To make contact to present experiments on a
quantitative level, we also combine SMFT with an LDA approach and obtain the condensate fraction in the presence of an external trapping potential.
\end{abstract}


\maketitle

\section{Introduction}
\label{sec:int}
The interplay between disorder and interactions in Bose systems has been a
vital field of research in condensed matter both in theory and
experiment. The line of investigation was mainly initiated by the
seminal work of Fisher et al.\cite{fisher-89prb546}, who first provided a detailed study  of localization of interacting bosons in a random
potential, which led to the notion of the superfluid-insulator
transition and the Bose glass (BG). While disorder effects in Fermi systems are relevant to a broad range of experimentally accessible scenarios
like correlated electron systems, the status was less diverse for Bose
systems for a considerable time period, as superfluid ${}^4$He
situated in random pores of Vycor had been the predominant setup which could be studies with sufficient precision~\cite{chan-88prl1950}.  This
changed dramatically when the realization of the superfluid-Mott
insulator transition of ultracold bosonic atoms in an optical
lattice opened up a new field of
investigation~\cite{greiner-02n39,jaksch-98prl3108}. In particular,
optical lattices provide a relatively pure and tunable simulation of
effective models used to describe solid state systems\cite{Hofstetter_fermionic_SF}, where effects
like disorder can also be realized in a controlled manner.  While several
alternative realizations of disorder in optical lattices, such as
multichromatic lattices with non-commensurate wavelengths\cite{Inguscio_non_interacting,Deissler_localization_quasiperiodic, Roth_Inguscio_two_color_lattice, Fallani_two_color_lattice} or
multi-species gases with strongly differing tunneling
rates~\cite{Castin_binary_disorder, ospelkaus-06prl180403} have been proposed recently,
speckle laser patterns are probably by now one of the most
efficient methods to establish disorder in cold atoms~\cite{lye-05prl070401,Aspect_speckle, aspect_1d_disorder, DeMarco_3D, Pasienski-arXiv0908.1182, Krueger_supression_BG}.  Therein, it is possible to combine the
speckle beam with the remaining apparatus of the optical lattice to
simulate disordered lattice systems with a high tuning accuracy and without other
side effects.

A variety of theoretical approaches~\cite{Lewenstein_Disorder_review} has by now been applied to
the disordered Bose Hubbard model (BHM), first introduced
for ultracold atoms by Jaksch et al.~\cite{jaksch-98prl3108},
which is described by the Hamiltonian
\begin{eqnarray}
\mathcal{H}_{\mbox{\tiny BH}}&=&-J \sum_{\ev{i,j}} ( b_i^\dag b_j^{\phantom{\dag}} +  \mbox{h.c.})\nonumber \\
&&  + \sum_i (\epsilon_i-\mu)n_i + \frac{U}{2}\sum_{i} n_i(n_i-1),\label{BH_Hamiltonian}
\end{eqnarray}
where $b_i^{\phantom{\dag}}$ ($b_i^{{\dag}}$) annihilates (creates) a
particle in the lowest band Wannier state at site $i$, $n_i={b}_i^\dag b_i^{\phantom{\dag}}$ is the local particle number operator, $J$ denotes the nearest
neighbor hopping energy in the lowest band, and $\mu$ is the chemical
potential. $\epsilon_i$ is an on-site energy shift, which in
our case is a spatially uncorrelated random variable drawn from a distribution
$p(\epsilon)$ and $U$ is the on-site repulsive interaction. The subscript $\ev{i,j}$ indicates the sum over all neighboring pairs of sites. Unless stated otherwise, we use the unit $U=1$.

Several quantum phases can exist within this model, such as the Mott insulator
(MI), the Bose glass, the condensed phase, commonly referred to
as the superfluid (SF), as well as the normal phase at finite
temperature. The transitions between these phases,
which constitute some of the first experimentally feasible quantum
phase transitions in bosonic systems, have attracted much attention.
Numerically, a powerful approach is Quantum Monte Carlo~\cite{krauth-91prl2307, Rieger_QMC, Haas_QMC, Zimanyi_QMC,prokofjev-04prl015703,Dang_QMC,Pollet_absence_direct_trans,Gurarie_phase_diag} (QMC). While MI and SF phase can be characterized
efficiently~\cite{krauth-91prl2307}, the BG phase and the vicinities
of the transition lines are significantly more complicated to be
adequately described. The main reason is that for finite size
calculations in general, it is problematic to capture the correct
description of the phase borders, which are essentially dominated by
rare events, which is also a problem for QMC methods. In most cases, exact
diagonalization studies are simply inadequate due to limited size and
number of particles, which often obscures essential physics
(however, there may appear aspects that can indeed be suitably
captured by small clusters~\cite{luhmann-07cm}). On the other hand, with a similar range of treatable system sizes as QMC, density matrix renormalization group (DMRG) is an efficient complementary method~\cite{rapsch-99epl559}. However, DMRG is currently only applicable in one spatial dimension and thus does not allow for a
description of effects in higher dimensional
lattices.  Analytically, renormalization group
analysis~\cite{singh-91prb3002, Phillips_missing_moment, Krueger_supression_BG}, slave boson theory~\cite{Dickerscheid_finite_T}, the strong coupling approach~\cite{Freericks-96prb2691}, and various different kinds of mean
field theory descendants have been applied to the BHM with and without
disorder~\cite{Rokhsar-91prb10328, krauth-92prb3137,sheshadri-93epl257,Sheshadri_RPA,krutitsky-06njp187,bissbort-cm08, Demarco_Mott_Melting, Zamponi_BH_Bethe}.
Arithmetically averaged mean-field theories, on the other hand, are incapable of resolving the BG phase at all for $T=0$~\cite{krutitsky-06njp187}, but also for $T>0$ impose an unphysically strong phase coherence, leading to an overestimation of the SF phase.
Other methods like the random phase approximation (RPA) which has
successfully been applied to the system without
disorder~\cite{sheshadri-93epl257, Sheshadri_RPA} again suffer from finite size
effects for the BHM with disorder, as the absence of translational
symmetry constrains its applicability to much smaller system sizes.

To circumvent this type of problems in describing the different phases
of the disordered BHM, we use the SMFT which has been previously introduced and applied to the disordered BHM
at zero temperature~\cite{bissbort-cm08}. There, it was
found that the method efficiently describes localization,
 is both valid in high dimensions and in the thermodynamic limit, capturing rare events with their respective statistical weight and includes dimensional effects. In particular, it was found that at fixed $\mu$, there exists a critical hopping strength,
above which the system remains superfluid for arbitrarily strong
disorder.

In this article, we present the SMFT in detail and investigate
how the results found for the disordered BHM at zero temperature
are modified at finite temperature. 
In addition to prototypical box disorder, we consider exponential speckle disorder  
to better simulate systems realized in current experiments. Here, the results are qualitatively different in the sense that for any finite disorder strength the MI  gives way to the BG.

The paper is organized as follows: In Sec.~\ref{sec:method}, the SMFT is
explained in detail. First the general scope is outlined, followed
by the definition of quantities computed within SMFT, such as
compressibility, local Greens functions and condensate fraction. In
Sec.~\ref{sec:phases}, the essential results for the disordered BHM at
$T=0$ are given, followed by the extension of the SMFT calculations to $T\ne 0$.  
In Sec.~\ref{sec:box} we extend the results presented for box disorder~\cite{bissbort-cm08}, including finite temperature effects.
Alternative types of disorder, in particular disorder induced by speckle lasers
are discussed for the BHM in 
Sec.~\ref{sec:speckle}.  As another possible source of disorder, we
discuss the effect of kinetic (hopping) disorder in the BHM in
Sec.~\ref{sec:hopping}. However, we find no sensible dependence of
the system on this parameter. The applicability of these results to current experiments essentially relies on the estimate
of experimental temperature, which is discussed on in Sec.~\ref{sec:temp}.
We find that the experimentally realized temperatures are
still far above the regime for which we resolve the previously stated
interesting phenomena, such as disorder-induced condensation and reentrant behavior.
LDA + SMFT calculations are discussed in Sec.~\ref{sec:lda} to provide a closer connection to experimentally measurable quantities.  In Sec.~\ref{sec:conc}, we conclude that the SMFT is an efficient theoretical approach to disordered Bose systems and promises an adequate
description of ongoing experiments.

\section{Method}
\label{sec:method}
As pointed out previously\cite{fisher-89prb546, krutitsky-06njp187}, performing a
self-consistent disorder average over all local on-site energies is
not sufficient to generally describe the insulating Bose-glass phase. From
spatially resolved bosonic Gutzwiller calculations, it becomes
apparent that this method overestimates long range correlations,
predicting the formation of a global condensate into a single particle
orbital which consists of the superposition of a large (extensive)
number of distinct localized single particle states. In the true Bose
glass phase, off-diagonal long range order does not prevail and a
large (although not extensive) number of particles may occupy each of
these localized modes independently, leading to a condensate fraction
$f_c=0$ in the thermodynamic limit. However, by imposing phase rigidity
and averaging over all mean-field parameters $\psi_i=\ev{b_i}$ with
the same complex phase, the spatially resolved (as well as the
arithmetically averaged) Gutzwiller theory leads to a finite average
mean-field parameter (MFP) and a finite condensate fraction
$f_c=\overline{\ev{b}}^2/\overline{\ev{b^\dag b}}$ in the expected BG
regime.

In contrast to the approaches mentioned above, the SMFT is
constructed as a single-site theory in the thermodynamic limit, which
effectively describes fluctuations in the MFPs using a
probability density function (PDF) $P(\psi)$. Set up in this fashion,
the SMFT method is not restricted to incorporating disorder
fluctuations only, but may also be a powerful approach
to treat fluctuations of different origins, such as of thermal or
quantum type in a unified framework. The concept of approximating quantum mechanical operators by random variables has been applied previously in a variety of physical scenarios~\cite{Yukalov_collective, Yukalov_spin_relaxation, Yukalov_PRA75}.
 In this paper, we will focus on disorder-induced
fluctuations at finite temperature. In this section, we will discuss
the construction of SMFT in detail, concentrating on the case of including on-site 
disorder. Extensions of including disorder in the hopping parameter $J$ or thermal fluctuations explicitly
are discussed in Sec.~\ref{sec:hopping} and App.~\ref{sec:explicit_thermal_fluctuations} respectively.

The central quantity, which effectively describes a disordered bosonic
lattice gas in the thermodynamic limit is the probability distribution $P(\psi)$.
It is assumed to be equal for all sites and in particular independent of the
nearest neighbors' on-site energies. The validity
of this assumption has been checked using spatially resolved mean-field
theory, which is known to become exact in the non-interacting limit,
while retaining interaction effects beyond Gross-Pitaevskii theory for
finite $U/J$. It is found that it yields correlation coefficients
below $0.05$ in all regimes considered\cite{bissbort-cm08}, justifying the above assumption.
The aim of SMFT is to find a self-consistent solution for
$P(\psi)$, which is restricted and uniquely specified by self-consistency
equations and minimization of energy.

Consider a cluster of lattice sites composed of a central site $i$ and $Z$ (the coordination number) nearest neighbors, where the corresponding part of the Hamiltonian~\eqref{BH_Hamiltonian} contains the operators on site $i$ within the bosonic Gutzwiller approximation~\cite{fisher-89prb546,Rokhsar-91prb10328,sheshadri-93epl257, krauth-92prb3137}:
\begin{equation}
\begin{split} \label{eff_ss_hamiltonian}
	\mathcal{H}_i^{\mbox{\tiny (MF)}}= -J \sum_{\mbox{\scriptsize n.n.}j}(\psi_j^* b_i^{\phantom{\dag}} + \psi_j^{\phantom{\dag}} b_i^\dag -\psi_j^*\psi_i^{\phantom{*}} )  \\
	   +(\epsilon_i-\mu) {b}_i^\dag b_i^{\phantom{\dag}} + \frac{U}{2}  b_i^\dag b_i^\dag b_i^{\phantom{\dag}}  b_i^{\phantom{\dag}}.
	   \end{split}
\end{equation}
Within this approximation, the site $i$ is coupled to (and the sum extends over) the nearest
neighbors via the scalar MFPs $\psi_j=\ev{b_j}$, which are random
variables within SMFT. Due to global particle number conservation, the MFPs can all be chosen to
be real and positive, as any variation in complex phase corresponds to a boost in local kinetic energy. The expectation value $\ev{b_j}$ is to be evaluated in the local ground state for $T=0$ or taking the thermal trace for $T>0$. Thermal fluctuations can, however, also be incorporated on an explicit stochastic level in the distribution $P(\psi)$ within SMFT, as discussed in App.~\ref{sec:explicit_thermal_fluctuations}.
Inspecting the Hamiltonian (\ref{eff_ss_hamiltonian}), it does not depend on the $\psi_j$'s individually, but only on
the scaled sum
\begin{equation}\label{eta_def}
\eta_i = J\sum_{\mbox{\scriptsize n.n.}j} \psi_j.
\end{equation}
Since the $\psi_j$'s (and in the hopping disorder Sec.~\ref{sec:hopping} also $J$)
are random variables, the newly defined quantity $\eta_i$ is also a
random variable obeying some distribution $Q( \eta)$.  As discussed
above, the random variables $\psi_j$ are assumed independent within SMFT,
allowing us to express the distribution $Q(\eta)$ by a $Z$-fold scaled
convolution
\begin{eqnarray}
Q(\eta)&=&\int_0^\infty d\psi_1 \, P(\psi_1) \ldots \int_0^\infty d\psi_Z \, P(\psi_Z) \;\nonumber \\
&&\times \, \delta \Bigl( \eta- J\sum_{m=1}^Z \psi_m \Bigr).
\label{def_Q}
\end{eqnarray} 
Making use of the convolution theorem, this can be reduced to two
one-dimensional Fourier transforms by introducing the characteristic
function
\begin{equation}
\varphi(t)= \int d\psi \, P(\psi) \, e^{it\psi},
\end{equation}
in terms of which the function $Q$ can be expressed as
\begin{equation}
	Q (\eta)=\frac{1}{2\pi J} \int dt \left[ \varphi(t) \right]^Z \, e^{-i t \eta / J} \label{Qdistribution}.
\end{equation}

This can be calculated efficiently for arbitrary coordination numbers
$Z$ using the FFT algorithm. The numerical procedure is discussed in
App.~\ref{multi_grid_fft}.

The self-consistency condition can now be formulated in the following
way: If the on-site energy $\epsilon$ is randomly drawn from
$p(\epsilon)$ and $\psi_j$ is randomly drawn from the
self-consistently determined distribution $P(\psi)$ for each of the
$Z$ nearest neighbors, this defines the single site Hamiltonian
(\ref{eff_ss_hamiltonian}), which can be diagonalized providing the ground
state $\ket{\mbox{g.s.}(\epsilon, \eta)}$. From there, a new MFP
$\bra{\mbox{g.s.}(\epsilon, \eta)}b \ket{\mbox{g.s.}(\epsilon, \eta)}$
can be calculated, with the self-consistency requiring the
distribution of this new random variable to be exactly the distribution
$P(\psi)$ we initially assumed for the neighboring $\psi_j$'s.

To cast this condition into functional form for a probability
distribution in the thermodynamic limit (i.e. for an infinitely large
system at fixed density), we first define the function

\begin{equation} \label{Def_g}
	g(\mu-\epsilon, \eta)=\bra{\mbox{g.s.}(\mu-\epsilon, \eta)}b \ket{\mbox{g.s.}(\mu-\epsilon, \eta)},
\end{equation}
where $\ket{\mbox{g.s.}(\mu-\epsilon, \eta)}$ is the ground state of $\mathcal{H}_i^{\mbox{\tiny
    (MF)}}(\mu-\epsilon, \eta)$ given in~(\ref{eff_ss_hamiltonian}).

It is useful to introduce the \textit{conditional} PDF, which is a
function of $\eta$ and $\psi$, giving the probability density for a
specified value $\psi$ if the value of $\eta$ is fixed and $\epsilon$
is distributed according to $p(\epsilon)$. This can be obtained by
using the transformation property of a PDF under a variable transform
and takes the form
\begin{align}\label{conditional_pdf}
\begin{split}
	\tilde P(\psi|\eta)&= \sum_{i\, \left|\, g(\mu-\epsilon_i,\eta)=\psi\right.}  \left|  \left( \frac{\partial g(\mu',\eta)}{ \partial \mu'} \right)_{\mu'=\mu-\epsilon_i} \right|^{-1} \; p(\epsilon_i)\\
	&=\frac{d}{d\psi} \int d\epsilon \: p(\epsilon) \: \Theta\left( \psi - g(\mu-\epsilon,\eta) \right).
\end{split}
\end{align}
This function does not obey any self-consistency condition and can be
evaluated directly, which is the first step in finding $P(\psi)$.
Remembering that the relation between $Q(\eta)$ and $P(\psi)$ is given
by (\ref{def_Q}), so that we can express the self-consistency condition as
\begin{align}
\label{SelfConsistency}
P(\psi)=\int d\eta\,  Q (\eta) \, \tilde P(\psi|\eta).
\end{align}
The right hand side can be understood as follows: for every fixed
$\eta$ we have a PDF $P(\psi|\eta)$, which yields a contribution with
the respective weight $Q(\eta)$, leading to a marginal distribution
for the considered site. If this agrees with the initially assumed distribution $P(\psi)$, a self consistent solution has been found.

Once this is determined, expectation values of local, self-averaging operators $\hat A$ can be directly be determined by
\begin{equation}
	\ev {\hat A}=\Tr {\varrho(\beta)},
\end{equation}
where
\begin{equation}
\begin{split}
	\varrho(\beta)=&\int d\epsilon\, p(\epsilon )\int d\eta \, Q(\eta) \frac{  e^{-\beta \mathcal{H}_i^{\mbox{\tiny
    (MF)}}(\mu-\epsilon,\eta)}}{\Tr{e^{-\beta \mathcal{H}_i^{\mbox{\tiny
    (MF)}}(\mu-\epsilon,\eta)}}}
\end{split}
\end{equation}
is an effective disorder averaged density operator, incorporating thermal, on-site energy and MFP fluctuations, depending explicitly on $p(\epsilon)$ and the self-consistently determined $Q(\eta)$.

\subsection{Numerical solution}
\label{sec:numerical_sol}
To solve the SMFT equations numerically, we iterate the
self-consistency equations on a discretized grid for $\psi$,
consisting of a superposition of a variable number of equidistantly
spaced grids, as explained in App.~\ref{multigrid}. For every fixed set
of physical parameters, we first numerically determine the conditional
cumulative density distribution function $F(\psi|\eta)=\int_0^\psi
d\psi' \, P(\psi'|\eta)$ for all values of $\eta$ and $\psi$ which
constitute the numerical grids for $Q(\eta)$ and $P(\psi)$ respectively
(discussed in App.~\ref{app:numerical_CDF}). Working with the cumulative
distribution on a numerical level, as opposed to the PDF itself, is
far more controlled and circumvents divergences in the PDF $P(\psi)$,
but also in the conditional PDF $P(\psi|\eta)$. The self-consistency
condition (\ref{SelfConsistency}) is not influenced by this approach.
As can be seen by inspection, the insulating solution
$P(\epsilon)=\delta(\epsilon)$ is always a self-consistent solution,
equivalent to the $\psi=0$ solution in the single site theory.
However, in the SF regime, there also exists a second, non-trivial
self-consistent solution, which corresponds to a lower grand canonical
potential and is therefore the physical solution in this case.
Furthermore, the physical solution is always found to be the
attractive fixed point of the self-consistency mapping in the space of probability distributions, i.e. if the iteration procedure is started at
any $P(\psi)\neq \delta(\psi)$, the successive distributions
continuously converge towards the physical distribution.

We start the iterative procedure with an initial PDF $P^{(0)}(\psi)$,
where all the weight is distributed at small, but non-zero values of $\psi$, assuring
fast convergence in the insulating state and in the vicinity of the
phase border. The distribution in the $i$-th iteration step for the scaled sum
of MFPs from the nearest neighboring sites $Q^{(i)}(\eta)$ is
calculated from $P^{(i)}(\psi)$ using the convolution theorem for
independent random variables and the FFT algorithm (see
App.~\ref{multigrid}). The new distribution $P^{(i+1)}(\psi)$ is then obtained by
integration over $\eta$. Numerically this is
done by using the trapezoidal rule, as we found that using higher order
techniques, such as Simpson's rule, is not robust and lead to
incorrect results if $\delta$-peaks appear in the PDF.

In the vicinity of the phase border, the computational effort
increases due to two effects. Firstly, the convergence is critically
slowed down, increasing the required number of iterations. Secondly, the
discretization of the $\psi$-grid plays an ever increasing role and in
some parameter regimes directly on the outside of the the MI lobes, where the
converged distributions are very close to a $\delta$-peak at $\psi=0$, the
numerically determined form of the distributions depends on the
discretization (resolution), which is clearly unphysical. In these
cases, we have to determine $P(\psi)$ by examining a sequence of
converged distributions at ever increasing resolution and define the physical
distribution as the limit of this sequence.

\section{Phases of the disordered BHM}
\label{sec:phases}

\subsection{Phases at $T=0$}
For the disordered BHM, three different phases exist at zero
temperature: 
A Mott insulating state, where
number fluctuations are suppressed and the particles are localized due
to a repulsive interaction. This state exhibits a finite energy gap of
order $U$, thus the single-particle density of states (DOS) at $\omega=0$ vanishes and the state is
incompressible.

If tunneling-induced delocalization dominates, the system is in a condensate 
(SF) phase, where a macroscopic number of particles can lower their energy by 
condensing into one single-particle state, thus exhibiting
quantum coherence and leading to a finite condensate fraction. Within
a grand-canonical mean-field description, this phase breaks the
$U(1)$-symmetry of the BH Hamiltonian (\ref{BH_Hamiltonian}) and leads
to a non-zero order parameter. The phase border from the SF to any of the 
insulating phases is thus determined by SMFT, where finite weight in $P(\psi)$ 
moves to finite values of $\psi$.

Finally, there is the Bose glass phase, where particles are localized by an 
interplay of disorder and interactions. However, there exists no single particle 
state which is occupied macroscopically and thus the BG is not a condensate, i.e. the
condensate fraction vanishes ($f_c=0$). However, there does exist an
extensive number of localized single particle states, each of which is
occupied by an arbitrarily large, but not macroscopic number of
particles. This may be understood as a highly fragmented system of
\textsl{incoherent} localized `non-macroscopic quasi-condensates'.

The transition from a BG to a MI is not determined from the self-consistent 
distribution $P(\psi)$, but by analyzing the compressibility $\kappa$ or the single particle DOS.

We consider the latter quantity within two frameworks:
\begin{enumerate}
	\item Considering the purely sing site particle- and hole excitations as specified 
	in \cite{krutitsky-06njp187}. The BG extends over the region where $P(\psi)=\delta(\psi)$ and values of the chemical potential $\mu \in \left\{m+\epsilon \,| \, m\in \mathbb{N}, \epsilon \in \mathbb{R}, p(\epsilon)>0 \right\}$, i.e. the borders are independent of $J$ and a direct MI-SF transition is possible.
	
	\item A more detailed analysis presented in \cite{Freericks-96prb2691,Pollet_absence_direct_trans} 
	relies on the analysis of an effective Hamiltonian in the subspaces of localized single particle- 
	and hole excitations. For finite hopping $J$ in the pure system, these hybridize, lifting the 
	degeneracy and form superpositions with quantum numbers $k$. Increasing (decreasing) $\mu/U$, 
	the Mott insulating state remains the ground state until the energy difference between the
	 MI state and the $k=0$ particle (hole) state vanishes, at which point particles delocalize 
	 and condense into these states. Now let us return to the disordered case: here the local 
	 particle (hole) excitations will not hybridize into fully delocalized states with well-defined 
	 quasi-momentum $k$, but into inhomogeneous states, which depend on the individual disorder configuration. 
	 It is however possible to make exact statements about the eigen-energy spectrum for a disordered system in the thermodynamic limit: The lowest kinetic energy is obtained in locally homogeneous regions and approaches the energy of the $k=0$ particle (hole) state of the pure system, as the size of this locally homogeneous Lifshitz region increases. Furthermore, scaling predicts that the dependence of the kinetic energy on the specific boundaries to the Lifshitz region will reduce, as the size of the region increases. On the other hand, the potential energy of the particle (hole) state is  minimized, when the local on-site energy takes on the lowest (highest) possible value over the whole region, i.e. lies at the extrema of $p(\epsilon)$. Therefore, upper (lower) phase boundary of the Mott lobe in the disordered system is obtained by shifting the upper (lower) boundary  down (up) by $\mbox{max}(\{\epsilon| p(\epsilon)>0\})$ 
	 ($\mbox{min}(\{\epsilon| p(\epsilon)>0\})$). The MI for box disorder in 3 dimension at $T=0$ 
	 obtained by strong coupling theory using this criterion, is the area enclosed by the orange phase boundaries in Fig.~\ref{box_phase_diags_T}. The insulating region outside of these lobes, bounded from the SF by the dashed white lines, corresponds to the BG. Using this criterion, the transition from MI to SF always occurs through the BG phase for box disorder within SMFT.

\end{enumerate}

\subsection{Phases at $T>0$}
At finite temperature $T>0$, the system is always compressible and the
incompressible Mott insulator is replaced by a normal (non-superfluid) phase with a
thermally induced compressibility. A central statement in \cite{krutitsky-06njp187} is, that the disorder-averaged single particle DOS calculated by considering purely local excitations only, allows for a clear distinction at $T>0$ between the BG and the normal phase, as it is zero in the latter. However, when considering delocalized excitations in the particle and hole sector, this statement holds no longer, as there are generally degenerate states in the $N$-particle particle-hole band and the particle (hole) band in the $N+1$ sector (or the $N-1$ sector for the hole band), which can be seen from the single particle DOS in the Lehmann representation
\begin{align}\label{eq_SPDOS_interact_finite_T_res}
\begin{split}
	\rho(\omega)= &\frac{1}{Z_{c}} \sum_{l,l',k} e^{-\beta E_{l}^{(N)}} \, \left[   |\varphi_{l'}^{(l,k)} |^2 \; \delta \left( \omega-(E_{l'}^{(N+1)} - E_{l}^{(N)}) \right)  \right.\\ & \left.+ |\gamma_{l'}^{(l,k)}|^2 \; \delta \left( \omega-(E_{l}^{(N)} - E_{l'}^{(N-1)})  \right) \right].
\end{split}
\end{align}

\begin{figure}[h!]
\label{Fig_Energy_Spec}
\begin{center}	\includegraphics[width=\linewidth]{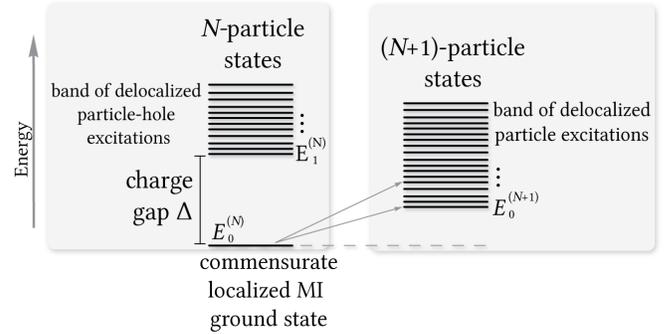}
\vspace{-6mm}
\end{center}
\caption{Energy structure of the BHM for large $U/J$ without disorder. The levels correspond to the energies of the exact many-particle energy eigenstates of the system with interactions. Due to the $U(1)$-symmetry the energy eigenstates can be chosen to be of well-defined particle number. For commensurate particle number $N$, the ground state (MI) is separated from a band of particle-hole excitations by a charge gap, while for $N+1$ particles, all eigenstates are delocalized, forming a band of hybridized particle states. At $T=0$, the single particle DOS has finite weight at frequencies corresponding to energy differences $E_{i}^{(N+1)}-E_{0}^{(N)}$ of transitions between $N$ and $N+1$ states. At $T>0$ transitions between different excited states also contribute, causing the gap to vanish if the two bands overlap and the respective matrix elements do not vanish.}
\end{figure}

Here, $E_{l}^{(N)}$ denotes the energy of the $l^{\mbox{th}}$ many-particle eigenstate $\ket{\psi_{l}^{(N)}}$ of $H_{BH}$, which can always be chosen to have well-defined particle number $N$, since $[\mathcal{H}_{\mbox{\tiny BH}}, \hat N]=0$, while $l$ is a label for the eigenstate in this subspace. $\varphi_{l'}^{(l,k)}=\bra{\psi_{l'}^{(N+1)}} b_k^\dag\ket{\psi_{l}^{(N)}}$ and $\gamma{l'}^{(l,k)}=\bra{\psi_{l'}^{(N+1)}} b_k\ket{\psi_{l}^{(N)}}$ are the amplitudes for the various possible transitions.
If the system is in a Mott insulating state (considering no disorder for the clarification of this argument) as shown in Fig.~\ref{Fig_Energy_Spec}, the ground state with $N$ particles is separated from a band of particle-hole excited states by a gap $\delta(J,U)$ (corresponding to the height of the Mott lobe at the respective $J/U$). As transitions between various excited states can also occur at finite temperature, the SPDOS thus vanishes if the particle-hole band with $N$ particles, overlaps with the particle- or hole band containing $N+1$ or $N-1$ particles, but the weight is suppressed exponentially by $\delta/T$. We thus conjecture, that the BG and the normal phase are not fundamentally distinguishable, and only connected by a crossover.

Following along the lines of  \cite{krutitsky-06njp187}, we use the disorder averaged local single particle density of states (DOS) at zero frequency $\overline \rho(\omega=0)$
(see Eq.~\eqref{DOS_fct_speckle}, \eqref{DOS_fct_box}), to determined the normal phase / BG  crossover at finite $T$. In the presence of disorder the SPDOS containing localized excitations can be calculated explicitly from the local Green's function in this regime (see App.~\ref{app_loc_DOS}) and leads to
\begin{equation}
\begin{split}
	\overline \rho&(\omega,\mu,\Delta,\beta)= \int d\epsilon \, p(\epsilon) \, \frac 1 {Z(\mu-\epsilon)} \\ 
	\times & \sum_{m=0}^\infty  e^{-\beta(\frac{U}{2}m(m-1)-\mu m +\epsilon m)}\left[ (m+1) 
  \delta(\omega-Um+\mu-\epsilon) \right. \\
  +& \left. m \, \delta(\omega-U(m-1)+\mu-\epsilon)  \right].
\end{split}
\end{equation}
$Z(\mu')$ is the local partition function at an effective chemical
potential $\mu'$ and the two $\delta$-distributions correspond to local particle and hole excitations respectively. Using this criterion leads to an overestimation of the MI / normal phase over the BG and thereby an absence of the BG phase around the tips of the Mott lobes.

\subsection{Deviations in Finite Size Systems}
Although phase transitions are, strictly speaking, only well defined in the thermodynamic limit, crossovers observed in current experiments may indicate phase borders that do not coincide with the borders obtained in systems of infinite size. As discussed above and in\cite{Freericks-96prb2691,Pollet_absence_direct_trans}, the MI/BG phase borders in an infinitely large  disordered system simply correspond to the shifted phase borders of the pure system. This argument relies on the existence of arbitrarily large Lifshitz regions, which is clearly no longer given for finite systems. The phase diagram for a finite system therefore strongly depends on the specific disorder realization, with the critical values for the phase borders becoming random variables. Therefore, a better question to ask for a finite system for instance is: For a randomly chosen disorder realization in a system consisting of $L$ sites, what is the probability $P_g$ that the energy gap will be lower than a given value $D$?
In the limit of $J/U\to 0$ this probability is given by
\begin{equation}
	P_g(\mbox{gap}<D)=1-(1-P_D)^L
\end{equation}
with the restriction that $0<D<U/2$ and
\begin{equation}
	P_D=\sum_{m=0}^\infty \int_{mU-D}^{mU+D} d\mu' \, p(\mu-\mu').
\end{equation}
With increasing $J/U$ the MI/BG phase border in the $J/U$-$\mu/U$-diagram remains a random variable, but with a reduced steepness in the slope for finite systems, as shown in \cite{Freericks-96prb2691, Zimanyi_QMC}. The BG/SF would also be very likely (in a statistical sense) to move to larger critical values of $J$ in a finite system, as the occurrence probability of `rare events', favoring a SF, is suppressed.

\section{Box Disorder}
\label{sec:box}
Results for box disorder at $T=0$ obtained by SMFT have been presented in
\cite{bissbort-cm08}, here we extend the phase diagram by taking collective excitation in the BG into account and focus on finite temperature. All numerical results presented in this paper are for a three-dimensional cubic lattice (Z=6).
 Box disorder is characterized by a constant probability density
for the on-site energies $p(\epsilon)=\frac{1}{\Delta}
\Theta(\Delta/2-|\epsilon|)$ over a bounded interval of width $\Delta$,
where $\Theta(x)$ denotes the Heaviside function. Changing the disorder
strength may have an influence on the system properties on various
levels. Coming from the insulating state within a local mean-field
picture, where the total state of the system is a direct product of
local Fock states minimizing the total energy, an increase in disorder
reduces the smallest possible particle or hole excitation energy. The
local energy gap for a site with an effective chemical potential
$\mu'$ and a Fock state as ground state $\ket{n=\max(\lceil\mu'/U
  \rceil,0)}$ is $E_{\mbox{\tiny particle}}=\max(\lceil\mu'/U
\rceil,0)\,U-\mu'$ for adding a particle and $E_{\mbox{\tiny
    hole}}=\mu'-(\max(\lceil\mu'/U \rceil,0)-1)U$ for creating a hole, if
$\mu'>0$. In the presence of disorder in an infinitely large system,
this implies that the energy gap is necessarily reduced by $\Delta/2$
and vanishes in the $J\to 0$ limit as soon as the interval of realizable effective chemical
potentials contains a positive integer multiple of $U$, i.e. $ \mathbb
N \, \cap [(\mu-\Delta/2)/U,\, (\mu+\Delta/2)/U ] \neq \emptyset$.

\begin{figure}[b!]
\begin{center}	\includegraphics[width=0.8\linewidth]{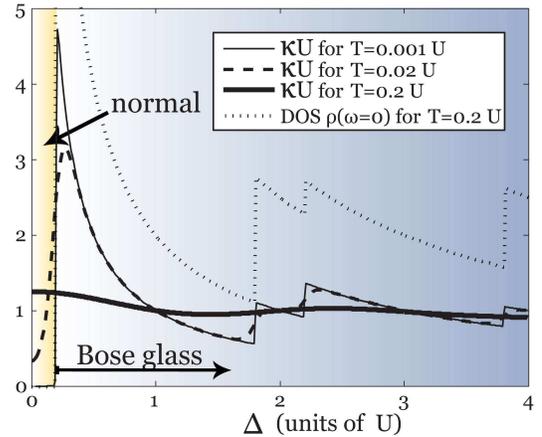}
\vspace{-6mm}
\end{center}
\caption{\label{fig_compressibility_of_Delta_box}(Color online) Compressibility $\kappa$ and local single particle DOS in an insulator for
  box disorder at $\mu=1.1U$ at various temperatures (in units of
  $U^{-1}$). At zero temperature $\kappa$ vanishes in the MI, but not
  in the BG and may thus be taken as a quantity to distinguish between them.
  At finite temperature however, $\kappa$ becomes
  non-zero in the normal phase and the sharp features become rounded
  off by thermal fluctuations. In this case, the disorder averaged DOS of local excitations $\overline{\rho}(\omega=0)$, which retains all sharp features at $T>0$, can be
  used to determine the crossover from a BG to the normal phase, where it remains
  zero.}
\end{figure}

Within the picture of purely local excitations, this argument is independent of temperature and directly reflected by the disorder averaged local DOS
\begin{equation}
\begin{split}
\label{DO_av_DOS_box}
\overline \rho(\omega,\mu,\Delta,\beta)=\frac{1}{\Delta}\sum_{m=0}^{\infty}(m+1)\frac{\Theta(\frac \Delta 2 -|\omega-Um+\mu|)}{ Z(Um-\omega)}  \\ \times \left[ e^{-\beta E_m(Um-\omega)}+ e^{-\beta E_{m+1}(Um-\omega)}\right].	\end{split}
\end{equation}

In the case of bounded disorder, this directly implies that the system
cannot be in a gapped state as soon as the carrier width ($\Delta$ for
box disorder) equals or exceeds $U$  and that the system is then always in the BG or SF state, i.e. the MI and normal state can only exist at $\Delta<U$
(and not at all for unbounded disorder). When the system is either of these two
insulating states at zero temperature, the compressibility
$\kappa=\frac{\partial \overline n}{\partial \mu}$ is determined by $\overline \rho(\omega=0)$ within a picture of local excitations, where locally degenerate sites
change their occupation number by $\pm 1$ if $\mu$ is slightly altered.
Since for finite temperature thermal particle or hole excitations are also present when the system is gapped, $\kappa$ is always driven to a finite value. The behavior of $\kappa$ and $\overline \rho(\omega=0)$ is
shown in Fig.~\ref{fig_compressibility_of_Delta_box} at the specific
chemical potential of $\mu=1.1 U$. Starting at $\Delta=0$, the system
is in a gapped MI / normal state, but increasing $\Delta$, the system
undergoes a phase transition at a critical disorder strength of
$\Delta=0.2U$.

At this point the local DOS $\overline \rho(\omega=0)$ takes on a
non-zero value, as two different local Fock states can become
degenerate at the edge of the box disorder distribution. At low
temperatures the compressibility exhibits a significant change from an
exponentially small value in $T$ in the normal phase to a large value
due to the existence of local configurations of on-site energies
which give rise to almost degenerate states with different particle
number. There, a small change in the chemical potential leads to a
local jump in particle number. In the zero temperature limit the
compressibility locally behaves as $\kappa \propto \frac {1}{\Delta}$,
since with increasing disorder strength the statistical weight of
these events is reduced. When the effective chemical potentials at the borders of the probability distribution enter a
region corresponding to a new local particle number $\mu \pm \frac
\Delta 2$, the compressibility increases with a jump, corresponding to
subsidiary phase transitions between different BG phases at $T=0$,
which turn into less pronounced crossovers with increasing $T$. 
Considering the limit of strong disorder $\Delta/U \to \infty$ at
fixed $\mu/U$ in Fig.~\ref{fig_compressibility_of_Delta_box}, the
system approaches a state in which only half the number of sites are
occupied and $\lim_{\Delta \to \infty} \kappa =\frac 1 {2U}$.

However, disorder does not only affect local excitation
properties, but leads to an intricate interplay between the hopping
and interaction energy scale, influencing the overall coherence
properties of the system. Whereas an increase in temperature or
interaction energy generally tends to counteract the formation of a
condensate, this is true in most, but not all scenarios when
increasing the disorder strength. In certain parameter regimes at sufficiently low
temperature, an increase in $\Delta$ can actually lead to the formation and stabilization of a condensate (i.e. disorder induced condensation), as previously predicted by various methods, including SMFT $\cite{bissbort-cm08}$.

\begin{figure}[t!]
\begin{center}
	\includegraphics[width=\linewidth]{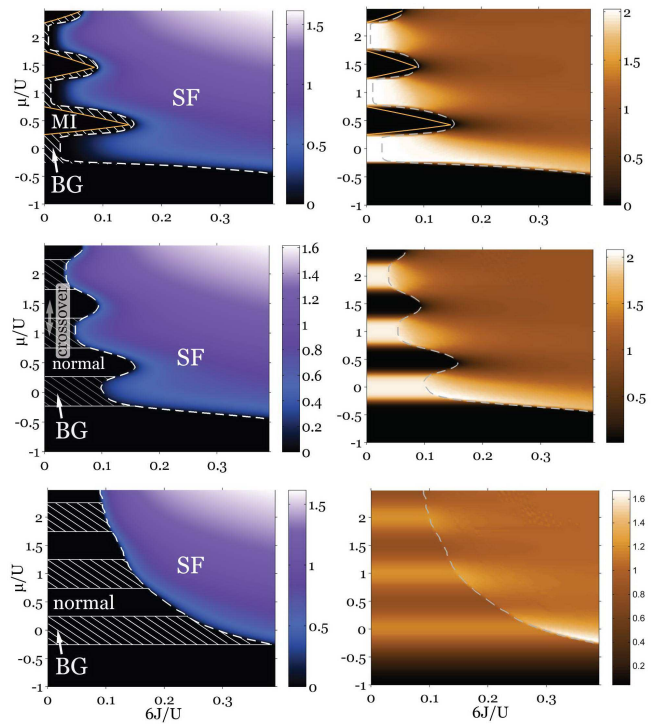}
\vspace{-6mm}
\end{center}
\caption{\label{box_phase_diags_T}(Color online) Finite temperature phase diagrams 
for box disorder with $\Delta=0.5 U$ for three temperature regimes:
  Zero temperature (upper row), intermediate $T=0.03U$ (middle row)
  and high temperature regime $T=0.2U$ (lower row). The left column
  shows the expectation value of the self-consistently determined
  distribution, which is the order parameter for the SF-insulator
  transition. It vanishes in the MI, BG, and the $T>0$ normal phase.
  In the $T=0$ diagrams, the MI/BG phase borders from strong coupling theory are indicated by the orange solid lines. The BG phase is marked by white stripes, at finite $T>0$ this is determined by  a vanishing local DOS only (i.e. not by strong coupling theory).
  At finite temperature there is only a crossover between the normal and BG phase, and the borders determined by the DOS of purely local excitations is marked by the horizontal white lines. The right column shows the compressibility $\kappa$ in units of $U^{-1}$. Whereas this is a suitable quantity to distinguish the MI (incompressible) from the BG at $T=0$, it is non-zero, but
  exponentially small in $T$ in the MI in the intermediate regime and of order
  $U^{-1}$ in the high temperature regime. The white borders indicate the transition to the SF, determined from the respective diagram on the left.}
\end{figure}

In Fig.~\ref{box_phase_diags_T}, the effect of increasing temperature
is exemplified in three phase diagrams and the compressibility in the
three main regimes: the zero, low and high temperature regime at fixed
disorder strength. The tips of the MI /normal lobes remain almost unchanged under an increase in $T$, while the BG region between the initial lobes is strongly enhanced and stabilized, even at a very low
temperature of $kT=0.03U$ (central plots of Fig.~\ref{box_phase_diags_T}). Since the SF/insulator phase border still
possesses a distinctive lobe structure, but the BG region is strongly
enhanced in contrast to the T=0 case (upper plots of Fig.~\ref{box_phase_diags_T}), we therefore refer to this situation as the
intermediate temperature regime. Furthermore, a large value of compressibility
is still a strong indicator for the BG at this
temperature, suggesting that it may still useful as an indicator of the 
transition between the BG and the normal phase in experiment.
In the high temperature regime $kT \gtrapprox
0.2U$, on the other hand, (see lower diagrams in Fig.~\ref{box_phase_diags_T}) the
compressibility is large throughout, approaching the value $\kappa \to U^{-1}$ 
in the high $T$ limit. The typical temperature of $T\approx 0.2U$ at which the MI 
and normal phase melts is consistent with previous previous studies \cite{Gerbier_finite_T}.  
In this  temperature regime the lobe structure of the SF/insulator phase border is totally wiped out and the critical border to the SF follows a $(JZ)_c \propto \mu^{-1}$ decay with increasing filling.

The SF/BG transition is highly sensitive to the system size. When the
system approaches the SF phase from the BG phase, the localized single
particle orbitals occupied by a large number of bosons increase in
size and are occupied by an ever increasing, but never extensive
number of particles in the BG phase. At the transition point the
localization length, being a measure for the size of these orbitals,
diverges and driven by percolation, phase coherence between
neighboring orbitals is established, eventually driving the system
into the SF phase. In finite size systems, the detection of the
transition point thus critically depends on the system size, as the BG
phase may be mistaken for the SF phase if the localization length is
larger than the system size. SMFT has the advantage that it is
constructed in the thermodynamic limit for an infinitely large system
in the grand canonical ensemble and takes all possible disorder
realizations into account within a functional description for the
probability distributions. The numerical error in the discretization
performed for the distribution $P(\psi)$ is well controlled and 
to be distinguished from the finite size deviations  made in real space calculations.

\begin{figure}[b]
\begin{center}
	\includegraphics[width=\linewidth]{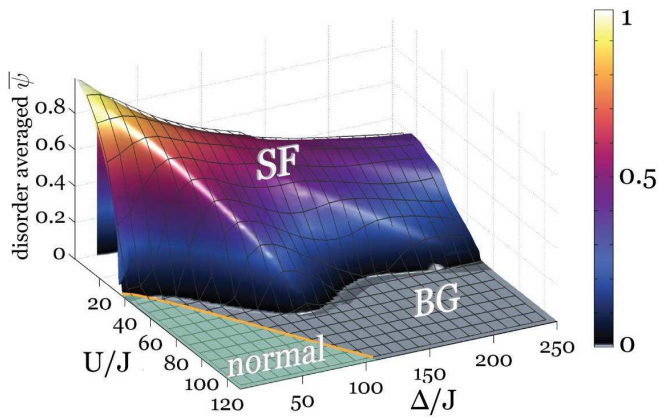}
\vspace{-6mm}
\end{center}
\caption{\label{box_phase_diags_n1}(Color online) Phase diagram for box disorder at fixed density $n=1$ in the $U/J$-$\Delta/J$ plane at $T=0.03U$ showing the mean order parameter $\overline \psi =\sqrt{f_c}$. Reentrant superfluidity is reflected by the protruding SF lobe (we also find a second less pronounced lobe at higher $\Delta$), where increasing $\Delta$ can drive the system through a sequence of SF-insulator transitions.
The orange line specifies the crossover from the normal to the BG phase (at $T=0$ this line becomes the MI-BG phase border) as determined by by shifted mean-field phase borders at this temperature, giving a better approximation than the strong-coupling approach for three dimensions. The simpler criterion of looking at purely local particle and hole excitations Eq.~\ref{DO_av_DOS_box} would lead to the MI/BG phase border at $U=\Delta$ and predict a direct MI-SF transition at small $\Delta$. Including collective excitations in the BG, as done here, always leads to an intermediate BG phase between the MI and SF in the $T=0$ limit.}
\end{figure}

In the phase diagram for box disorder Fig.~\ref{box_phase_diags_T}, 
SMFT does not give rise to a direct transition from the MI to the SF, if the 
extended criterion for the MI/BG border, including collective excitations in rare regions is used.
This is furthermore demonstrated in the finite temperature phase diagram at 
constant filling $n=1$ in Fig.~\ref{box_phase_diags_n1}. Due to the absence of a clear 
distinction between the normal and BG phase at $T>0$, the orange border only indicates a crossover between these regimes, but would go over into a MI/BG phase border for $T=0$. At any $\Delta>0$ a finite BG region intervenes between the MI and SF phases.

The question whether this transition always occurs via the BG phase has been a highly debated topic since the
introduction of the disordered BHM \cite{fisher-89prb546}, and was
established for the one and two dimensional case
\cite{rapsch-99epl559,Svistunov_RG_1D,prokofjev-04prl015703}. In a
recent work \cite{Pollet_absence_direct_trans} it was shown that this
scenario is true in any finite dimension for bounded disorder, due to
the statistical certainty that any possible configuration of on-site
energies for a cluster of sites will occur in the limit of an
infinitely large system. In the previous work \cite{bissbort-cm08} these 
collective excitations in the BG phase were not considered and within a simpler framework
of purely local particle and hole excitations, a direct transition was predicted. 

It is known that arithmetically  averaged MFT, as well as SMFT providing an improvement in any finite dimension, both become exact in the limit of infinite dimensions, where no BG exists at $T=0$. However, as argued recently\cite{Pollet_absence_direct_trans} the theorem of inclusions guarantees the existence of an intervening BG phase between the SF and the MI in any arbitrarily high, but finite number of dimensions for bounded disorder. It is instructive to understand the decrease of the BG region, including the collective excitation in the Lifshitz regions in terms of percolation physics: For any finite dimension the outer border between the BG and SF phase specifies the critical value $(J/U)_{\mbox{\tiny crit.1}}$ specifies the lowest energy at which it becomes energetically favorable for the particles to form a global condensate (in SMFT this is the border where $\overline \psi$ takes on a finite value). The border between the MI and BG inside the global insulator, specifies the critical value $(J/U)_{\mbox{\tiny crit.2}}$ at which it becomes possible for the system to form large local superfluid patches (locally resembling pure systems) without phase coherence between different patches. With increasing dimensionality of the system, the connectivity between different patches increases (percolation is enhanced) and the required tunneling energy $(J/U)_{\mbox{\tiny crit.1}}$ to form one large percolated patch, i.e. a global condensate, decreases. In the limit of high dimensions, the critical values of $J$ at which these two phenomena occur approach the arithmetically averaged mean-field value\cite{krutitsky-06njp187}
	\begin{align}
\begin{split}
	JZ_c(\mu) =&  \Delta \left[n \ln \left( \frac{1-n+\mu + \Delta/2 }{1-n+\mu - \Delta/2 } \right) \right.\\
	&+\left.(n+1) \ln \left( \frac{n-\mu + \Delta/2 }{n-\mu - \Delta/2 } \right) \right]^{-1},
\end{split}
\end{align}
and the BG disappears, where $n$ is the filling and $\mu$ and $\Delta$ are given in units of $U$.

In Fig.~\ref{box_phase_diags_n1} we present a phase diagram calculated at fixed density $n=1$ in the low temperature regime $T=0.03U$. At every point in the diagram, the self-consistent distribution is calculated for a fixed $\mu$, enabling the calculation of the density $\ev{n(\mu,\Delta,U,J,T)}$. Thereafter $\mu$ is iteratively determined using Ridder's algorithm~\cite{numerical_recipes} until the density obtained from SMFT does not deviate more than $\Delta n=0.005$ from the specified density.
In Fig.~\ref{box_phase_diags_n1}, the disorder averaged MFP $\overline \psi =\int \psi \, P(\psi) d\psi$ (within SMFT, this is exactly $\sqrt{f_c n}$) clearly shows the usual SF/insulator phase transition (at fixed low temperature) along the line $\Delta=0$, where the disorder localizes the particles with increasing $U/J$. Moving outwards into the $\Delta/J$ at fixed interaction $U/J$, the condensed phase is surprisingly robust, surviving local on-site fluctuations $\Delta$ several hundred times larger than the hopping energy $J$, as pointed out in a recent work\cite{Gurarie_phase_diag}. This can be understood from the bosons filling up the low-lying sites and forming a 'background sea' via the repulsive interactions, creating an effective smoother potential in which it is energetically favorable for the remaining bosons to delocalize\cite{bissbort-cm08}.
A remarkable effect at sufficiently low temperature is the appearance of a SF lobe, protruding into the insulating domain at finite $\Delta$. In this regime the interplay between disorder and interactions is non-monotonic in these two effects, and, for the regime $40\lessapprox U/J \lessapprox 85$, an increase in $\Delta$  drives the system into the SF phase, delocalizing the particles. This effect can be understood from the pure BHM $\mu/U$-$J/U$ phase diagram and relies on the existence of a lobe structure, i.e. requires a sufficiently low $T$. To keep the particle number constant with increasing $\Delta$, $\mu$ is required to increase. In certain regimes the majority of sites in the system may enter from an insulating regime between the lobes into a regime inside the lobes, thereby favoring condensation.
Qualitatively, the SMFT phase diagram agrees well and shows the disorder induced SF lobe, as found in recent QMC calculation~\cite{Gurarie_phase_diag} at $T=0$ on relatively small lattices ($L=8\times8\times8$). 

At large $\Delta$ the order parameter is non-monotonic in $U/J$, vanishing at sufficiently small $U/J$ which indicates a transition into an Anderson localized state, where the localization almost exclusively disorder-induced. However, the region of extremely small $U/J$ is problematic as $\Delta/U$ and $\mu/U$ diverge, since very few sites have to contain an ever increasing number of particles to keep the disorder-averaged density fixed, when asymptotically half the number sites (due to the symmetry of the box distribution $p(\epsilon)$) have such a high effective on-site energy, that they contain no particle. Due to the diverging local occupation number, this limit transcends the constraints imposed in the derivation of the BHM in an optical lattice and is, in this sense, unphysical.


\section{Speckle Disorder}
\label{sec:speckle}
Although a homogeneous box distribution is most commonly used for
disorder calculations in theory, it is currently not an experimentally
feasible choice. In this section we discuss and compare the results
for a realistic disorder distribution created by a speckle laser to those of a
box disorder distribution. A laser passing
through an inhomogeneous disordered plate leads to a disordered optical
potential, which is the Fourier transform of the disordered pattern on
the plate. In recent experiments, it has become possible to reduce the
autocorrelation length of this disordered potential to the order of the
lattice spacing ($\leq 1\mu m$) \cite{DeMarco_3D}. With this
experimental achievement, the priorly most criticized artifact of a
speckle laser for creating uncorrelated disorder has been overcome,
thereby making speckle potentials the most promising method for future disorder
experiments in optical lattices.  The resulting distribution for uncorrelated
on-site energies is well approximated by
\begin{equation}
	p(\epsilon)=\frac{\Theta(\epsilon)}{\Delta} e^{-\epsilon/\Delta}
	\label{speckle_distribution}
\end{equation}
Although it may be argued that an optical speckle potential in experiment is fundamentally bounded by its finite size, it is only essential that the width of the on-site energy distribution exceeds $U$ (which is fulfilled in essentially all experimentally relevant regimes). This justifies the use of $(\ref{speckle_distribution})$.

\begin{figure}[t]
\begin{center}
	\includegraphics[width=\linewidth]{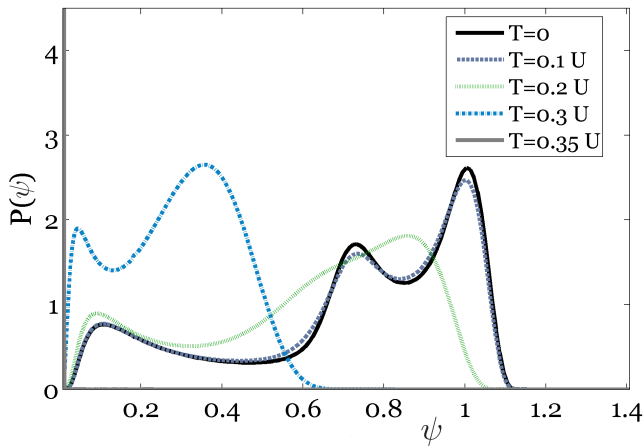}
\vspace{-6mm}
\end{center}
\caption{\label{fig:dists_speckle_various_T}(Color online) Typical self-consistent distributions $P(\psi)$ for speckle
  disorder for $\Delta=1U$, $\mu=1.2U$, $Z=6$, and $JZ=0.3U$ at different temperatures. For low temperatures the
  mean-field parameters in the condensed phase are robust against
  finite temperature fluctuations, but with increasing temperature the
  system is eventually driven into an insulating BG phase, as indicated by the distribution at $P(\psi)=\delta(\psi)$ in the at $T=0.35U$.}
\end{figure}

To treat this disorder distribution using SMFT, it is useful to
perform a transformation of variables $x(\epsilon)=-e^{-\epsilon/\Delta}$, which on a formal level transforms the SMFT
conditional probability functions into a form analogous to homogeneous
disorder. This step enters only on the level of calculating
the conditional cumulative distribution function (CDF),
\begin{equation}\label{speckle_cond_CDF}
	F(\psi|\eta)=\lim_{c\to 0}\int_{c-1}^{0} dx \, \Theta(\psi-g(\mu+\Delta \, \ln(-x),\eta)).
\end{equation}
Apart from this, the SMFT method remains identical to the homogeneous
disorder case. Similarly, arbitrary disorder distributions may also be
incorporated into SMFT, although an analytical transformation of the
random variable will not exist in general.

In contrast to box disorder, which has been the distribution primarily
focused on so far when considering the disordered BHM, speckle disorder is
unbounded and arbitrarily high values of $\epsilon$ have a finite
probability to occur. This leads to the  effect
(strictly only possible for an infinitely large system) that
\textit{turning on} the disorder by an arbitrarily small amount
immediately changes a large part of the phase diagram from the MI /
normal phase into the BG phase. 

In a physical picture, the effective chemical potential $\mu'$ can then take 
on integer multiple values of $U$ (the on-site potential can become 
arbitrarily high) with a non-zero
probability density, where local Fock states with different particle
number become degenerate, leading to a finite compressibility and
local DOS $\overline \rho(\omega=0)$ at $T=0$.

\begin{figure}[t]
\begin{center}
	\includegraphics[width=\linewidth]{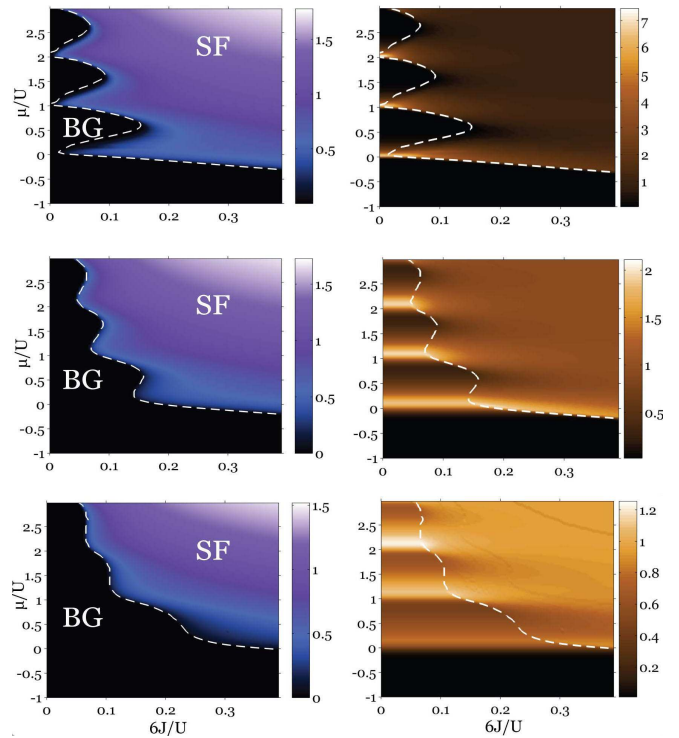}
\vspace{-6mm}
\end{center}
\caption{\label{fig:dists_speckle_various_Delta}(Color online) Disorder averaged MFP $\overline \psi = \int d\psi \, \psi \, P(\psi)$ characterizing the SF-insulator transition (left column), and compressibility (right column, in units of $U^{-1}$) in the
  $JZ/U$-$\mu/U$-plane for \emph{speckle disorder}. Diagrams are shown for increasing disorder strength ($\Delta=0.1 U$, $\Delta=0.3U$, $\Delta=1 U$) in  the zero or low temperature regime ($T=0$, $T=0.05U$, $T=0.05U$) for the (upper, middle, lower) row respectively.  In contrast to the behavior for
  box disorder, the structure of each lobe does not change
  symmetrically, but rapidly extends in the direction of decreasing
  $\mu/U$ with increasing disorder strength $\Delta$. Since the MI /
  normal phases do not exist for speckle disorder, the insulating
  region (black in the left figures) is always a BG. The white lines
  denote the SF / insulator phase boundaries, indicating where $\overline
  \psi$ takes on a finite value.}
\end{figure}

In the absence of disorder and at sufficiently high tunneling coupling $J$,
a macroscopic number of particles occupy the $\ket {k=0}$ Bloch state.
Within an effective, symmetry breaking Gutzwiller description the local order
parameters $\psi_l=\ev b_l$ then take on a finite and constant value,
reflecting the translational symmetry of the system. Within SMFT this
state is characterized by a $\delta$-distribution
$P(\psi)=\delta(\psi-\psi_0)$, where $\psi_0$ is the order parameter
of conventional bosonic Gutzwiller theory. Turning on the disorder in
such a system in the SF state breaks the translational symmetry of the
system, i.e. the condensate state deviates from the $k=0$ Bloch state,
which is reflected by the distribution of MFPs $P(\psi)$ taking on a
finite width. Initially for weak disorder, an increase in disorder
always leads to a broadening of $P(\psi)$, but for stronger values of
$\Delta$ the system may eventually be driven toward an insulating
state, driving $P(\psi) \to \delta(\psi)$ and thereby decreasing the
fluctuations in the MFPs.  On the other hand, increasing the temperature
suppresses the SF and leads to a decrease of the MFPs above a certain
temperature, up to which the SF remains stable, as shown in 
Fig.~\ref{fig:dists_speckle_various_T}.

\begin{figure}[b]
\begin{center}
	\includegraphics[width=\linewidth]{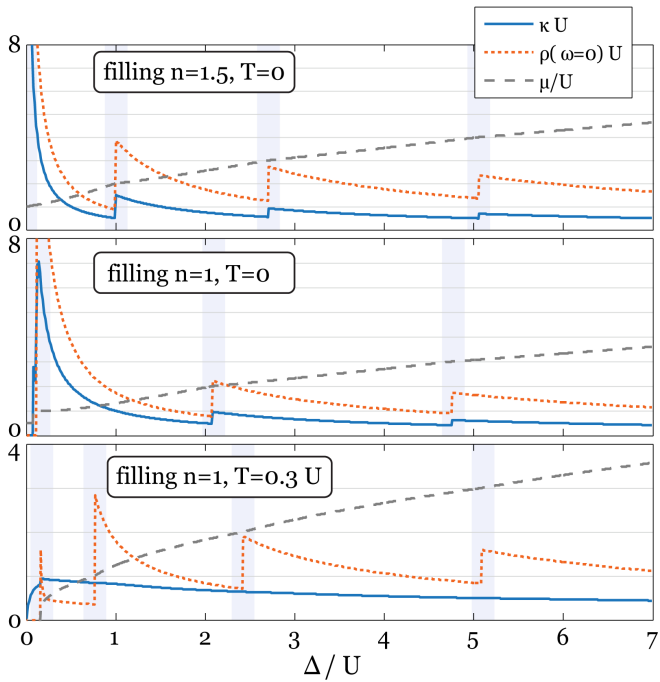}
\vspace{-6mm}
\end{center}
\caption{\label{fig:kappa_of_Delta_speckle}(Color online) The
  compressibility, local single particle DOS and chemical potential as a function of
  disorder strength $\Delta$ for speckle disorder at constant
  filling. Comparison of the upper two plots shows a diverging
  compressibility and local DOS for $\Delta \to 0$ for non-integer
  filling (the system remains superfluid), while for integer
  filling it drops to an exponentially small value in $\Delta$ at a
  value of $\Delta \approx 0.1U$ and vanishes in this limit. The
  lowest figure shows the same quantities at finite temperature
  $kT=0.3U$, where thermal fluctuations have totally smeared out the
  sharp features in the compressibility. However, these persist in the
  local DOS, although their position is changed due to a temperature
  induced shift in $\mu$.}
\end{figure}

The influence which speckle disorder has on the $\mu/U$-$J/U$-phase
diagrams is shown in Fig.~\ref{fig:dists_speckle_various_Delta}. In
contrast to box disorder, where the distribution of on-site energies
is symmetric around $\mu$ and the insulating lobes give way to the SF
in the same way on the upper and the lower side of the lobe, the insulator
forms on the lower side of the lobes with increasing $\Delta$ for
speckle disorder. This can be understood from the fact that only lower
values of the effective local chemical potential can occur.

For strong disorder $\Delta \gtrapprox U$, the lobe structure of the
insulator / SF phase boundary is washed out, which is similar to the
effect of finite temperature. For speckle disorder, $\kappa$
cannot be used to identify a phase transition, since it is non-zero in
both the BG and the SF. A question of interest, regarding the disappearance of the MI /normal
phase for an arbitrarily small amount of speckle disorder, is how the
compressibility behaves as a function of $\Delta$, as some interpretation 
is needed, that an 'infinitesimal amount of disorder' can
instantaneously convert the whole MI/normal area of the phase diagram
into a BG.

In Fig.~\ref{fig:kappa_of_Delta_speckle}, $\kappa(\Delta)$ is
shown for different parameters to clarify this dependence.  In the
insulating state, the compressibility can be calculated explicitly
(\ref{insulator_kappa}) and one obtains
\begin{align}\label{kappa_insulating_speckle}
\begin{split}
	\kappa&=\frac{1}{\Delta}  \left[  \frac{ \sum_m m\, e^{-\beta E_m(\mu)} }{\sum_m  e^{-\beta E_m(\mu)}} \right.\\
	&\left.- \frac 1 \Delta \int_0^\infty d\epsilon \,e^{-\frac \epsilon \Delta} \frac{ \sum_m m\, e^{-\beta E_m(\mu-\epsilon)} }{\sum_m  e^{-\beta E_m(\mu-\epsilon)}} \right]
\end{split}
\end{align}
Essentially, two different scenarios have to be considered. First, if
$\mu/U$ is positive and integer, $\kappa$ diverges in the limit of vanishing
disorder, as is well known from the pure BHM phase diagram, where the
density is a step function in $\mu/U$ for $J=0$. This is equivalent to the case
for non-integer, fixed particle density $n$, where the system remains SF for any non-zero $J$. Second, if $\mu/U$ is
non-integer, the compressibility vanishes with decreasing $\Delta$, as
the system approaches a point in a Mott lobe away from the border.
This corresponds to the case of fixed, integer-valued density $n$.

We will now discuss the behavior of $\kappa$ and $\overline\rho(\omega=0)$
 at fixed particle density $n$, shown in Fig.~\ref{fig:kappa_of_Delta_speckle}. Keeping the density constant
with rising disorder, requires the chemical potential to be
increased, as an ever increasing number of sites shifts to weights
with lower occupation numbers. At every point when $\mu/U$ passes a
positive integer number, a new Fock state becomes potentially
occupied, but with an ever decreasing statistical weight as $\Delta$
increases. As a result the compressibility experiences a jump at each of these points, as highlighted by the gray regions in Fig.~\ref{fig:kappa_of_Delta_speckle}, where $\mu/U$ (dotted gray lines in Fig.~\ref{fig:kappa_of_Delta_speckle}) passes an integer value. This leads to the characteristic series of ever smaller
kinks in $\kappa(\Delta)$. At finite temperature, these features in $\kappa$ are
smeared out over a typical scale of $kT$, whereas the sharp features in the local single particle DOS survive at $T>0$ within SMFT (solid blue lines in Fig.~\ref{fig:kappa_of_Delta_speckle}).
\begin{figure}[bth]
\begin{center}
	\includegraphics[width=\linewidth]{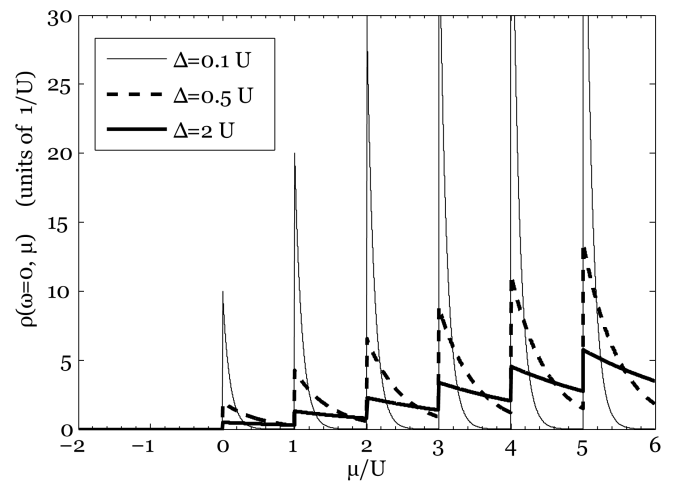}
\vspace{-6mm}
\end{center}
\caption{\label{fig:speckle_DOS_along_mu} The disorder averaged single particle density of states
  $\rho(\omega=0, \mu)$ at zero frequency for $T=0.05 U$ for different
  speckle disorder intensities. For $\Delta=0$ this consists of a
  sequence of $\delta$-peaks at positive integer values of $\mu/U$,
  however for any $\Delta>0$ this quantity is non-zero for $\mu\geq 0$
  and the system is in the BG phase.}
\end{figure}

To clarify the effect speckle disorder has on $\overline \rho(\omega=0)$ and the immediate disappearance of the MI / normal phase at any $\Delta>0$, the local DOS is plotted for weak ($\Delta=0.1U$), intermediate ($\Delta=0.5U$) and strong ($\Delta=2U$) in Fig.~\ref{fig:speckle_DOS_along_mu} as a function of $\mu$. In the pure system $\overline \rho(\omega=0,\mu)$ consists of a sum of $\delta$-peaks at integer values of $\mu/U$, i.e. at these values of the chemical potential there are two degenerate Fock states $\ket{n= \mu/U }$ and $\ket{n=\mu/U+1}$ at all sites in the insulator and the local single particle DOS diverges. As soon as speckle disorder is turned on, these $\delta$-peaks are broadened according to the on-site energy distribution (\ref{speckle_distribution}) and $\overline \rho(\omega=0)$ takes on the form of a sequence of superimposed exponential functions, each decaying with the constant $\Delta$. From this it is clear that $\overline \rho(\omega=0)$ takes on a non-zero value as soon as $\Delta>0$ at any $\mu>0$, although it is exponentially suppressed for most values of $\mu$ at weak disorder $\Delta\ll U$. At zero temperature the different amplitudes of the various peaks in $\overline \rho(\omega=0,\mu)$ at integer $\mu$ can be exclusively attributed to the $\sqrt{n}$ factor from the action of the bosonic operators, whereas at $T>0$ the amplitudes (but not the positions of the sharp features) may also be modified by the Boltzmann factors in (\ref{DOS_fct_speckle}).

\begin{figure}[h!]
\begin{center}
	\includegraphics[width=\linewidth]{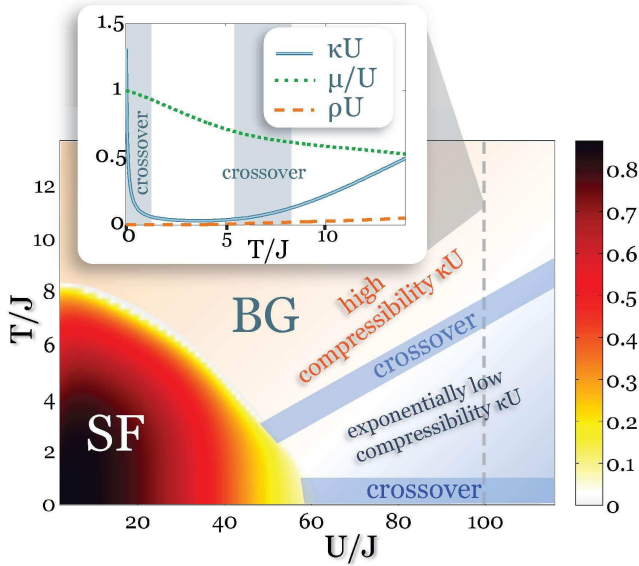}
\vspace{-6mm}
\end{center}
\caption{\label{fig:speckle_PD_T_U}(Color online) Main image: Phase diagram in the $T/J$-$U/J$-plane for fixed $\Delta=10 J$ at fixed filling $n=1$ in three dimensions. For this specific disorder strength the SF region is enlarged by the disorder in comparison to the pure case (i.e. disorder induced condensation, see Sec.~\ref{sec:reentrant_sf}).In the lower left corner, the SF exists at sufficiently low $T$ and $\Delta$ and the value of disorder-averaged $\overline \psi$ is color-coded. Outside the SF region, the system is always in a BG phase, but undergoes two crossovers from a regime with exponentially low compressibility $\kappa U$ for intermediately low $0.02 \lessapprox T/U \lessapprox 0.065$ (for $\Delta/J=10$) into a strongly compressible BG regime, in the limits of both high and very low $T/U$. To clarify the quantitative behavior, the compressibility $\kappa$, the chemical potential $\mu$ and the local single particle DOS $\rho(\omega=0)$ along the dashed line $U/J=100$ is shown in the inset.}
\end{figure}

Fig.~\ref{fig:speckle_PD_T_U} shows the phase diagram at constant
disorder strength $\Delta/J=1$ and constant density $n=1$ in units of
$J$. The SF region prevails in the lower left region at low temperature and weak interaction $U/J$. In this parameter regime the disorder stabilizes the SF phase, actually extending the SF region of the phase diagram in contrast to the pure ($\Delta=0$) case (disorder-induced condensation, see Sec.~\ref{sec:reentrant_sf}). All of the non-SF region in Fig.~\ref{fig:speckle_PD_T_U} is a BG, since we are dealing with unbounded disorder with $\Delta>0$, but we can identify a weakly and two strongly compressible regimes in the phase diagram. Since $P(\psi)=\delta(\psi)$ in the BG, the energy scale $J$ cannot influence thermodynamic quantities beyond a scaling relation, implying that the compressibility $\kappa U$ may be a function of $T/U$ only. Therefore the compressibility has a radial structure and it suffices to consider its behavior along a single line (such as $U/J=100$, depicted in the inset of Fig.~\ref{fig:speckle_PD_T_U}). This reveals that there are three regimes in this phase diagram: at high temperature $T/U\gtrapprox 0.065$ the system is strongly compressible ($\kappa U$ is of order unity), as thermal fluctuations have wiped out the sharp peaks over a wide range in $\mu$. At intermediately low temperature, the compressibility is exponentially low, as the typical thermal excitation energy scale does not suffice to excite the majority of sites to higher states. Somewhat surprisingly, within a certain parameter regime for $\Delta$ and the density $n$, the system undergoes a further crossover at very low temperature $T/U\approx 0.02$ into a second highly compressible regime. At $T=0$ and at integer filling, the chemical potential $\mu/U$ approaches an integer value from below (green dotted line in the inset of Fig.~\ref{fig:speckle_PD_T_U}). At these points $\kappa$ diverges in the pure limit of $\Delta \to 0$, which reveals that the compressibility is grows with the inverse of $\Delta$ and only persists in the limit $T \to 0$ if the density $n$ is integer.

\subsection{Disorder-induced reentrant Superfluidity}

\label{sec:reentrant_sf}
\begin{figure}[h!]
\begin{center}
	\includegraphics[width=\linewidth]{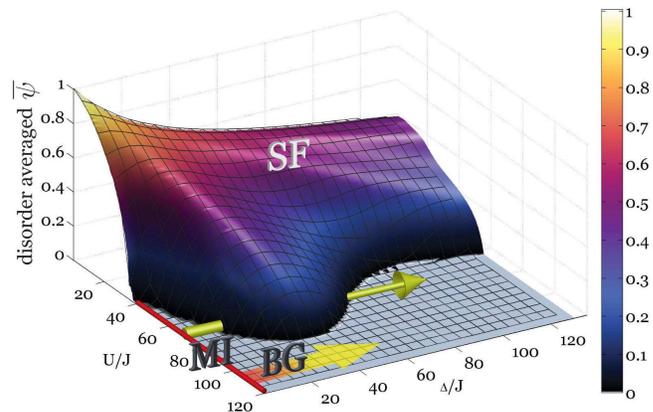}
\vspace{-6mm}
\end{center}
\caption{\label{fig:speckle_PD_Delta_U_low_T}(Color online) Low temperature $k T=0.03 U$ phase diagram showing the
  disorder-averaged MFP $\overline \psi$ for speckle disorder at
  constant filling $n=1$ in three dimensions. Multiple reentrant
  behavior can be seen within a small window of $U/J$. The red line at
  $\Delta=0$ indicates the presence of MI / normal state, for any
  $\Delta>0$ the insulator is a BG.}
\end{figure}

\begin{figure}[hb!]
\begin{center}
	\includegraphics[width=\linewidth]{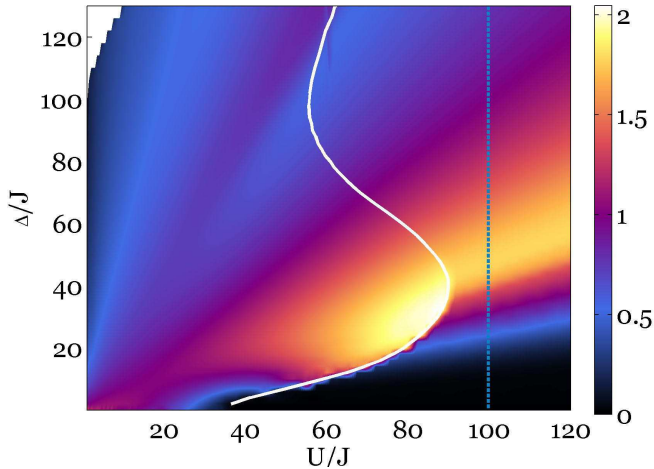}
\vspace{-6mm}
\end{center}
\caption{\label{fig:speckle_kappa_Delta_U_low_T}(Color online) Compressibility $\kappa$ in units of $1/U$ at $k T=0.03 U$ for speckle disorder at constant filling $n=1$ in three  dimensions (same parameters as in Fig.~\ref{fig:speckle_PD_Delta_U_low_T}). The incompressible normal phase
  only exists on the line $\Delta/J=0$, but the region where $\kappa$
  is exponentially small in the insulator extends linearly with $U/J$.
  The white line is the SF - insulator phase border, where $\overline
  \psi$ becomes zero. The compressibility along the blue dashed line can be understood as the low temperature case and compared to $\kappa$ in the lower two plots of Fig.~\ref{fig:kappa_of_Delta_speckle}, where the sharp features have been washed out by temperature, but are still pronounced peaks.}
\end{figure}

At low temperature and fixed density, we find that an increase in
$\Delta$ can actually drive the system from an insulating into a
condensed state within a certain window of $U/J$, as shown in
Fig.~\ref{fig:speckle_PD_Delta_U_low_T} and Fig.~\ref{fig:speckle_kappa_Delta_U_low_T}. Within a very small
window of $U/J$, the system may be driven through an additional
sequence of BG and SF phases.

On the $\Delta=0$ line, the usual SF - insulator transition occurs, whereas along $\Delta$ for small values of $U$, disorder only suppresses the SF slightly up to reasonably large values of $\Delta/J$. For even larger disorder
values, the system will eventually undergo the transition into an

Anderson localized state. In comparison to the corresponding phase diagram for box disorder Fig.~\ref{box_phase_diags_n1}, the extruding superfluid lobe appears at a considerably smaller values of $\Delta \approx 40J$ for speckle disorder ($\Delta \approx 100J$ for box disorder, not the different $\Delta$-axis scale). However, it should be noted that the measure $\Delta$ is not the same (in a statistical sense) for the two disorder types: whereas for speckle disorder $\Delta$ is also the standard deviation (std), in the case of box disorder the std is only $\frac{\Delta}{2\sqrt 3}\approx 0.29\Delta$, corresponding to weaker disorder for the same $\Delta$. Examining the position of the superfluid lobe in units of the standard deviation, actually reveals that it appears at slightly smaller disorder strengths for box disorder.

\begin{figure}[htb!]
\begin{center}
	\includegraphics[width=\linewidth]{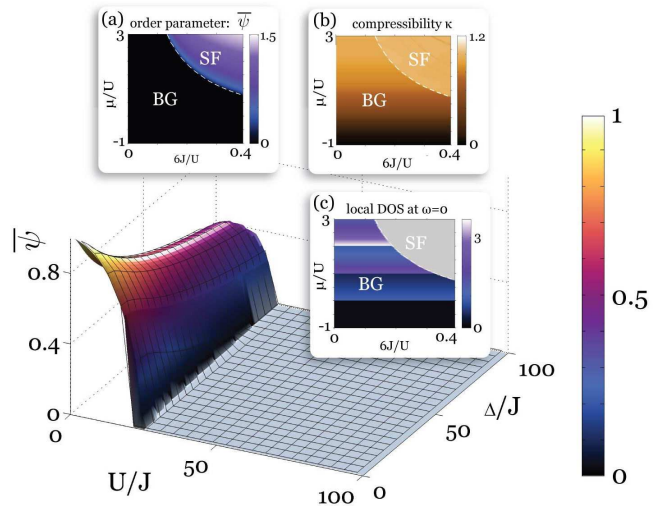}
\vspace{-6mm}
\end{center}
\caption{\label{fig:speckle_PD_Delta_U_high_T}(Color online) Main background image: same $\overline \psi$ phase diagram as in Fig.~\ref{fig:speckle_PD_Delta_U_low_T}, but in the experimentally relevant high temperature regime $kT=0.3U$ at $n=1$, where disorder cannot induce condensation.
  Insets: disorder averaged MFP (a), compressibility (b) and local DOS
  $\overline \rho(\omega=0)$ in the $\mu/U$-$JZ/U$-plane at the same temperature $T=0.3U$ for $\Delta=1 U$.
  In this regime most interesting structure has been washed out by
  thermal fluctuations and disorder and an increase in either $U$, $\Delta$ or $T$ \emph{always} suppresses condensation.}
\end{figure}

In the vicinity of the phase border, $\kappa$ is generally larger in
the SF state than in the neighboring insulator, as a shift in $\mu$
leads to a locally continuous shift in density. On the lower side of
the lobe at disorder strengths $\Delta\lessapprox 40J$, the
compressibility jumps across the phase border (i.e. second
order transition). At higher values of $\Delta$, $\kappa$ is almost
unaffected by the disappearance of the condensate, i.e. the
compressibility is almost exclusively induced by disorder. In the insulating state the ensemble of mean-field states determined by SMFT cannot depend on $J$ (since different sites are only coupled via $\psi$). Therefore any
physical quantity, such as $\kappa$, can only be a function of $T/U$
in this regime (reflected in the radial structure of $\kappa$ in the insulator).

In the high temperature regime relevant for current experiments (see App.~\ref{sec:temp} for an approximation of $T$),
the lobe in the $\Delta/J - U/J-$plane at fixed density has vanished completely and an
increase in disorder always counteracts the condensate formation. This
might explain the recent finding, that no reentrant behavior or
disorder-induced condensation was observed in  experiments so far~\cite{Pasienski-arXiv0908.1182}.  Furthermore, we
calculated the $\mu/U-J/U-$phase diagram for this temperature and
disorder regime, where the lobe structure is also fully washed out
and the system is dominated by thermal fluctuations, as seen in the insets of Fig.~\ref{fig:speckle_PD_Delta_U_high_T}. We therefore conclude that an upper critical temperature, which may depend on the filling, exists for the occurrence of reentrant superfluidity. Our temperature estimation suggests that $T$ is too high in current experiments to observe this effect, which agrees with recent experimental evidence~\cite{Pasienski-arXiv0908.1182}.

\subsection{Hopping Disorder}
\label{sec:hopping}
In addition to diagonal on-site energy disorder, it is possible to
incorporate off-diagonal hopping disorder~\cite{DellAnna_hopping_do} into the SMFT formalism.
In this case, the local hopping energy $J_{\ev{i,j}}$ between site $i$ and $j$ becomes a random 
variable, each described by a distribution $p_J(J)$, which we assume to be independent of the on-site energies $\epsilon_i$. This leads to the BH Hamiltonian with on-site, as well as hopping disorder
\begin{eqnarray}
\mathcal{H}_{\mbox{\tiny BH}}&=&- \sum_{\ev{i,j}} J_{\ev{i,j}}( b_i^\dag b_j^{\phantom{\dag}} +  \mbox{h.c.})\nonumber \\
&&  + \sum_i (\epsilon_i-\mu)n_i + \frac{U}{2}\sum_{i} n_i(n_i-1).\label{BH_Hamiltonian_hop}
\end{eqnarray}
The PDF $p_J(J)$ has been calculated for
a speckle disorder potential~\cite{Zhou-arXiv0907.5053, DeMarco_3D} using imaginary time evolution.

In contrast to diagonal disorder (on-site energy or interaction), where the
fluctuations are incorporated into the conditional PDF (\ref{conditional_pdf}) before the iteration procedure of the SMFT self-consistency equations, the hopping disorder acts as an additional source of fluctuations during this iteration procedure, methodically entering at a different point in the method.
The corresponding mean-field Hamiltonian $\mathcal{H}_i^{\mbox{\tiny (MF)}}$ , (analogous to (\ref{eff_ss_hamiltonian}))
now depends on the rest of the system only through the new random variable
\begin{equation}\label{eta_def}
\eta = \sum_{j=1}^Z J_{j} \psi_j,
\end{equation}
where both $J_{j}$ and $\psi_j$ are random variables, the latter again being assumed to be distributed according to $P(\psi)$. It is therefore convenient to introduce the an intermediate random variable $\phi=J \psi$, distributed according to the PDF $P_\phi(\phi)$. As explained in App.~\ref{PDF_of_products}, this new PDF can be expressed explicitly in terms of the PDF's  $P(\psi)$ and $p_J(J)$ as
\begin{equation}
\label{P_phi_conv}
	P_\phi(\phi)=\int dx \, p_J(e^x) \, P(\phi\cdot e^{-x}),
\end{equation}
numerically allowing the use of the FFT algorithm on a suitable grid. Once $P_\phi(\phi)$ is known, the PDF for the random variable $\eta=\sum_{l=1}^Z \phi_l$ can subsequently be calculated by
\begin{equation}
	Q (\eta)=\frac{1}{2\pi} \int dt \left[ \vartheta(t) \right]^Z \, e^{-i t \eta} \label{Qdistribution_hop},
\end{equation}
where 
\begin{equation}
\vartheta(t)= \int d\phi \, P_\phi(\phi) \, e^{it\phi}.\label{char_fct_hop}
\end{equation}
is the characteristic function of $P_\phi(\phi)$. For the numerical computation of the previous two, the FFT algorithm can be used.

The conditional PDF (\ref{conditional_pdf}) $\tilde P(\psi|\eta)$, incorporating the effect of any diagonal (here: on-site) disorder remains unchanged under the inclusion of hopping disorder and is calculated before the SMFT iteration procedure in full analogy to the previous case for the relevant on-site disorder type $p(\epsilon)$.
The final self-consistency equation, closing the iteration procedure is also left unchanged to the previous case (\ref{SelfConsistency})
\begin{align}
\label{SelfConsistency_hop}
P(\psi)=\int d\eta\,  Q (\eta) \, \tilde P(\psi|\eta),
\end{align}
except that the $Q(\eta)$ entering is calculated from (\ref{Qdistribution_hop}).

Of course the case of pure on-site disorder with a constant hopping energy $J_0$ can be obtained as a limit of this extension by setting $p_J(J)=\delta(J-J_0)$, causing (\ref{P_phi_conv}) and (\ref{Qdistribution_hop}) to reduce to (\ref{def_Q}).

The numerical iteration procedure is carried out analogously to Sec.~\ref{sec:numerical_sol}, except that only a single equidistant grid, as restricted by (\ref{P_phi_conv}), can be used. This limits the numerically obtained precision of $P(\psi)$ at small values of $\psi$.

In our calculation we use the appropriately scaled distribution $p_J(J)$, matching the on-site disorder strength, as obtained by Zhou and Ceperley\cite{Zhou-arXiv0907.5053} for a lattice depth of $s=14E_R$ and at a speckle disorder strength of $s_D=1E_R$, where we assume that the standard deviation in $p_J(J)$ is proportional to the disorder intensity $s_D$ (as motivated by their analysis). 
Since a change in the disorder strength does not change the most probable value of the distribution\cite{Zhou-arXiv0907.5053} $p_J(J)$, we model the hopping disorder distribution to have its most likely value at the given value $\overline J/U$, with the width being independent thereof. If any weight lies at values of negative $J$, this is set to zero and the distribution is subsequently renormalized, however this is only relevant for points deep in the insulator and does therefore not influence the results in any way. However, this only occurs at very small values of $\overline J/U$ deep in the insulator, making this formal alteration of $p_J(J)$ irrelevant to the result.

\begin{figure}[h!]
\begin{center}
	\includegraphics[width=\linewidth]{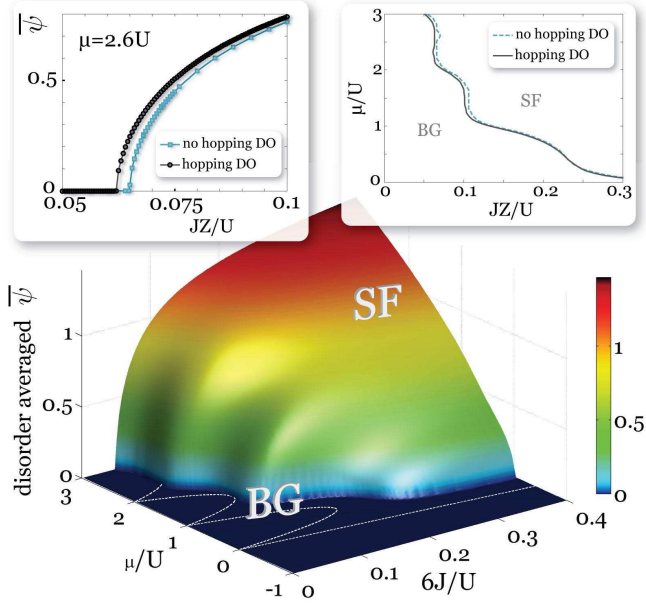}
\vspace{-6mm}
\end{center}
\caption{\label{fig:speckle_hopping_PD_JZ_mu}(Color online) Diagrams showing the effect of \emph{hopping disorder} at $T=0.05U$ and an on-site speckle disorder strength $\Delta=U$. Main 3D figure: disorder-averaged MFP $\psi$ for speckle on-site and experimentally corresponding hopping disorder. For orientation purposes, the MF phase boundary for the pure system is plotted as a thin dashed white line. The value $J/U$ refers to the most likely value of the hopping disorder distribution $p_J(J)$, whereas the width of the distribution $p_J(J)$ is constant. Upper left inset: comparison of the order parameter at fixed $\mu=2.6U$ with and without hopping disorder, clearly demonstrating that the addition of hopping disorder stabilizes the SF phase and leads to a lower critical value of $J$ in this regime. Upper right inset: Subsequent comparison of phase borders in the $\mu/U$-$J/U$-plane.}
\end{figure}

As shown in the low temperature phase diagram in the $J/U$-$\mu/U$-plane (Fig.\ref{fig:speckle_hopping_PD_JZ_mu}), the additional inclusion of hopping disorder leads to a stabilization of the SF phase and a small shift in the phase boundary. For disorder distributions in the typical experimentally relevant parameter regime, the standard deviation of the hopping parameter distribution $p_J(J)$ is three orders of magnitude smaller than the standard deviation of the on-site energy distribution $p(\epsilon)$ (as found in \cite{Zhou-arXiv0907.5053}, specifically for the distributions used here $\sigma_J/\sigma_\epsilon=0.0014$). This explains the minor, but clearly resolved modification of the phase boundaries in contrast to the pure on-site disordered case, shown in the upper right inset of Fig.~\ref{fig:speckle_hopping_PD_JZ_mu}.

\section{Incorporating experimental aspects}

\subsection{Temperature estimation in an optical lattice}
\label{sec:temp}
The initial temperature in the trap, prior to the optical lattice ramp-up, can be determined from the expansion profile of the cloud. If the
ramp-up of the optical lattice is performed adiabatically, the entropy
of the system is conserved and can actually lead to a cooling of the
atoms.  The initial entropy of a weakly
interacting cloud in the trap using Bogoliubov theory leads to
the expression~\cite{Rey_temperature}
\begin{equation}
	S_{\mbox{\tiny Bog.}}(\beta)=k_B \sum_{\mathbf p}\left( \frac{\beta \epsilon_{\mathbf p}}{e^{\beta \epsilon_{\mathbf p}}-1} -\ln[1-e^{\beta \epsilon_{\mathbf p}}] \right).
\end{equation}
After ramping the lattice to a sufficiently high intensity, the
entropy can be calculate up to first order (neglecting terms
containing $J$)~\cite{Rey_temperature}

\begin{equation}
	S_{\mbox{\tiny J=0}}=k_B\left[ -\beta \mu  + \frac 1 N \ln(\Xi(M))+\beta E\right],
\end{equation}
where $M$ is the number of sites and $\Xi(M)$ is the grand canonical
partition function.  Equating these two expressions for a sufficiently
high initial temperature $k_B T_{\mbox{\tiny initial}}>0.05 E_R$ leads
to the relation~\cite{Rey_temperature}
\begin{equation}
\label{final_temp}
	k_B T_{\mbox{\tiny final}} \approx \frac{U}{3E_R} (k_B T_{\mbox{\tiny initial}} +0.177 E_R)
\end{equation}
For $\phantom{|}^{87}$Rb in an optical with a wave length of $\lambda=812nm$ and intensity $s=11E_R$, the disorder-averaged interaction constant is $\overline U/E_R=0.355$. As a typical, conservative estimate of the initial temperature before the lattice ramp-up in current experiments~\cite{DeMarco_phase_slip} we use the value $T_{\mbox{\tiny initial}}=0.13 \mu K$, for which the relation (\ref{final_temp}) predicts a final temperature of $k_B T\approx 0.11E_R=0.32U$ after the ramp-up.

\subsection{LDA incorporating trap effects}
\label{sec:lda}
To compare the results obtained via SMFT to experimental data on a
quantitative level, we performed a LDA+SMFT calculation to
incorporate the effect of the trapping potential. Two effects of the lattice laser beams are
taken into account: The red-shifted lattice laser
beam with a Gaussian profile and beam width $w_0$ leads to an
attractive potential via the ac Stark effect. Furthermore, the local energies of states within
a localized Wannier basis are also increased in regions of high
intensity, leading to the renormalized effective lower trapping
frequency within an harmonic approximation \cite{Greiner_PhD}.
\begin{equation}
	\omega_{\mbox{\tiny eff}}^2= \frac{4 E_R}{m w_0^2} ( 2 s -\sqrt{s}).
\end{equation}

With the addition of an external magnetic trap with trapping frequency $\omega_{\mbox{\tiny mag}}$, the total trapping frequency, which the atoms are exposed to is given by

\begin{equation}
\label{total_trap_freq}
	\omega_{\mbox{\tiny tot}}=\sqrt{	\omega_{\mbox{\tiny eff}}^2+\omega_{\mbox{\tiny mag}}^2}.
\end{equation}

\begin{figure}[bth]
\begin{center}
		\includegraphics[width=\linewidth]{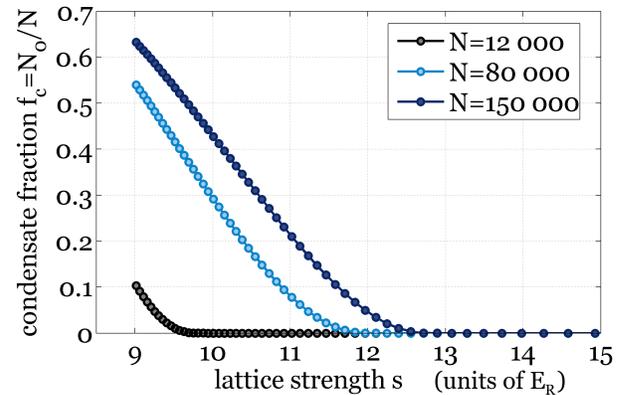}
\vspace{-6mm}
\end{center}
\caption{\label{fig:LDA_fc}(Color online) The condensate fraction as a function of the lattice strength
  $s$ for various total particle numbers in a harmonic trap,
  calculated within LDA and SMFT (connected dotted lines). The
  calculations were performed at fixed temperature $kT=0.3U$ and
  disorder strength $\Delta=1U$.}
\end{figure}

For every fixed value of the lattice height $s$, $J$ and $U$ are
extracted from the single particle Wannier function (beyond the
approximation for deep lattices~\cite{Greiner_PhD}), the total
trapping frequency (\ref{total_trap_freq}) is calculated and the local
effective trap energies are assigned on a sufficiently large $3D$
lattice. The chemical potential $\mu$ is adjusted using Ridder's
method~\cite{numerical_recipes} to obtain a specified total particle number. The condensate
fraction, depicted in Fig.~\ref{fig:LDA_fc}, is subsequently averaged
over the local values obtained by SMFT, weighted by the respective
density.

We used the following values for the experimental parameters: $w_0=110
\mu m$, the lattice laser wavelength was set to $\lambda_{\mbox{\tiny
    lat}}=812 nm$, and the magnetic trapping frequency
$\omega_{\mbox{\tiny mag}}=2 \pi \, 40Hz$.

In contrast to the behavior of the order parameter at the transition
point, the condensate fraction does not follow a power law decay in
the finite trap, but is smeared out a the transition point, as
superfluid regions in the trap decay in size with increasing $s$.

\section{Conclusion}
\label{sec:conc}
We have described the stochastic mean-field theory in detail on a
methodological level, and extend it to incorporate finite temperature
effects. Subsequently, we have applied it to ultracold atoms in an optical lattice with uncorrelated on-site box
disorder distribution and discussed the intricate interplay between
interaction, tunneling energy, disorder, filling and finite
temperature effects. Furthermore, we have presented, to the best of our knowledge, the first quantitative
theoretical calculations for speckle disorder, which leads to a qualitatively different
phase diagram than for box disorder and are of immediate experimental
relevance~\cite{Pasienski-arXiv0908.1182}. For this case, we have discussed
the characteristic features of the various phases and presented phase diagrams, both at fixed chemical potential and at fixed density.  Below a critical temperature, we find disorder-induced
condensation and multiple reentrant behavior, both for box and speckle disorder. The temperature in recent
experiments~\cite{Pasienski-arXiv0908.1182} is estimated and found to
be too high yet to observe disorder-induced condensation. We also find that including hopping disorder in addition to local on-site disorder for a realistic distribution of speckle parameters enhances the insulator and jumps in the order parameter within the SF phase indicate a series of transitions. An LDA+SMFT calculation has been
performed to incorporate the effects of an external trap for on-site speckle disorder.

\begin{acknowledgments}
  We thank I. Bloch, L. Carr, B. DeMarco, E. Demler, A. Pelster, D. Semmler, M. Snoek, H. Stoof and
  W. Zwerger for useful discussions. This work was supported by the
  German Science Foundation (DFG) via Forschergruppe FOR 801. UB and RT
  acknowledge support by the Studienstiftung des deutschen Volkes.
  Calculations were performed at the Center for Scientific Computing
  at the University of Frankfurt/Main and the TKM computation cluster
  at the University of Karlsruhe.
\end{acknowledgments}

\begin{appendix}
\section{Multi-Grid discretization}\label{multigrid}
\label{multi_grid_fft}
\begin{figure}[bth]
\begin{center}
	\includegraphics[width=0.5\linewidth]{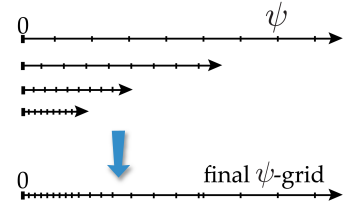}
\vspace{-6mm}
\end{center}
\caption{\label{fig:grids}Illustration of the multi-grid procedure.}
\end{figure}

To employ the FFT algorithm an equidistant grid is required, which is
not compatible with having a very high resolution at small values of
$\psi$ ($\sim 10^{-9}$) to capture the behavior in the vicinity of the
SF-insulator transition, as well as simultaneously correctly
describing the distribution at large values ($\psi \sim 1$). To
circumvent this problem, we use a superposition of equidistant grids,
which enables us to use the FFT algorithm on each of these grids
individually (see Fig.~\ref{fig:grids}).
This procedure relies on the following property of the convolution
(\ref{def_Q}): the $Z$-fold convolution of the truncated function
$P_t(\psi)=\Theta(a-\psi) \,P(\psi)$ with $P(\psi)=0$ for $\psi<0$
and $a<\psi_{\mbox{\tiny max}}/Z$ is identical to the $Z$-fold
convolution of $P(\psi)$ up to an easily determinable normalization
constant on the interval $[0,a]$. Here $\psi_{\mbox{\tiny max}}$ is
the largest value of the grid, if this is finite (as in the
discretized numerical case). In our calculations the number of grids
used (typically $\approx 1 \ldots 6$) with $200\ldots 1000$ points per
grid is adjusted dynamically within the iteration procedure, depending
on the position of the most likely value of $P(\psi)$ and the
convergence properties.

\section{Numerical calculation of the CDF $F(\psi|\eta)$ for box disorder}
\label{app:numerical_CDF}

We found the numerically most efficient method for calculating and tabulating the two-dimensional CDF $F(\psi|\eta)$ to be the following:

For every fixed value of $\eta$, consider the function (\ref{Def_g})
$g(\mu',\eta)$ on the interval $\mu'=(\mu-\epsilon) \in
[\mu-\Delta/2, \mu + \Delta/2]$. We define the local minima and maxima
of this function (including the end points) in increasing order as
$\{\mu_1^{\mbox{\tiny min}},\,\mu_2^{\mbox{\tiny min}},\ldots \}$ and
$\{\mu_1^{\mbox{\tiny max}},\,\mu_2^{\mbox{\tiny max}},\ldots \}$.
Furthermore the $n$-th monotonically increasing and decreasing
function on the restricted interval, provided that this interval
exists, is denoted by
\begin{align}
	g^{(\mbox{\tiny inc},n)}(\mu',\eta)&=g(\mu',\eta) \quad \mbox{for} \quad \mu'\in(\mu_n^{\mbox{\tiny min}}, \tilde \mu_n^{\mbox{\tiny max}})\\
	g^{(\mbox{\tiny dec},n)}(\mu',\eta)&=g(\mu',\eta) \quad \mbox{for} \quad \mu'\in(\mu_n^{\mbox{\tiny max}}, \tilde \mu_n^{\mbox{\tiny min}})
\end{align}
where
\begin{align}
	\tilde \mu_n^{\mbox{\tiny max}}&=\min_m \{ \mu_m^{\mbox{\tiny max}}| \mu_m^{\mbox{\tiny max}} > \mu_n^{\mbox{\tiny min}}   \}\\
	\tilde \mu_n^{\mbox{\tiny min}}&=\min_m \{ \mu_m^{\mbox{\tiny min}}| \mu_m^{\mbox{\tiny min}} > \mu_n^{\mbox{\tiny max}}   \}
\end{align}
By construction, these functions are invertible in $\mu'$ within the
defined range, allowing us to introduce the functions
\begin{widetext}
\begin{align}
	h^{(\mbox{\tiny inc},n)}(\psi,\eta)=\left\{
	\begin{tabular}{l l}
	$0$ &if $\psi \leq g(\mu_n^{\mbox{\tiny min}},\eta)$\\
	$\left(g^{(\mbox{\tiny inc},n)}\right)^{-1}(\psi,\eta) - \mu_n^{\mbox{\tiny min}}\quad$ &if $ \psi\in ( g(\mu_n^{\mbox{\tiny min}},\eta),\, g(\tilde \mu_n^{\mbox{\tiny max}},\eta)  )$\\
	$1$&if $\psi\geq g(\tilde \mu_n^{\mbox{\tiny max}},\eta) $	
   \end{tabular}
  \right.\\  
  	h^{(\mbox{\tiny dec},n)}(\psi,\eta)=\left\{
	\begin{tabular}{l l}
	$0$ &if $\psi \leq g(\tilde \mu_n^{\mbox{\tiny min}},\eta)$\\
	$\tilde \mu_n^{\mbox{\tiny min}}- \left(g^{(\mbox{\tiny dec},n)}\right)^{-1}(\psi,\eta) \quad$ &if $ \psi\in ( g(\tilde \mu_n^{\mbox{\tiny min}},\eta),\, g( \mu_n^{\mbox{\tiny max}},\eta)  )$\\
	$1$&if $\psi\geq g( \mu_n^{\mbox{\tiny max}},\eta) $	
   \end{tabular}
  \right.
\end{align}
\end{widetext}

in terms of which the CDF can be written as a superposition
\begin{equation}
	F_{\eta}(\psi)=\frac 1 \Delta \sum_n \left( h^{(\mbox{\tiny dec},n)}(\psi,\eta)+h^{(\mbox{\tiny inc},n)}(\psi,\eta) \right)
\end{equation}

\section{PDF for a product of random variables}
\label{PDF_of_products}
Including hopping disorder into the SMFT leads to the task of having
to calculate the probability distribution for the newly defined random
variable $\phi=J \psi$, if the PDFs for $J$ and $\psi$ are known
functions $p_J(J)$ and $P(\psi)$. This can be done by taking the
logarithm $s=\ln(\phi)=x+y$ and using the convolution theorem for the
PDFs $P_x(x)=e^x \, P_J(e^x)$ (analogously for $P_y(y)$) of the random
variables $x=\ln(J)$ and $y=\ln(\psi)$.  The resulting distribution
for $\phi$ can be denoted in compact form as
\begin{equation}
	P_\phi(\phi)=\int dx \, P_J(e^x) \, P(\phi\cdot e^{-x}).
\end{equation}
Using the FFT-algorithm for the numerical computation of this
operation leads to a vast increase in performance, however it requires
the functions $P_x(x)$ and $P_y(y)$ to be interpolated on an
equidistant grid.

\section{Compressibility in the insulating phases}
In the special case with $P(\psi)=\delta(\psi)$, the disorder
integration can be performed explicitly and expressions for the
density and the compressibility can be found. Using partial
integration and the property $\lim_{\epsilon \to \infty} p(\pm
\epsilon)=0$ for the on-site probability distribution, the disorder
averaged density at finite temperature can be expressed as
\begin{align}\label{kappa_insulating}
\begin{split}
	\overline n(\beta,U,\mu,\Delta)&=\int d\epsilon \,p(\epsilon) \, \frac{\sum_{m=0}^\infty m e^{-\beta E_m(U,\mu-\epsilon)}}{{\sum_{m=0}^\infty  e^{-\beta E_m(U,\mu-\epsilon)}}}\\ &=-\frac{1}{\beta}\int d\epsilon \,p(\epsilon) \, \frac {\partial}{\partial \epsilon} \left( \ln \sum_m e^{-\beta E_m(\mu-\epsilon)} \right)\\
	&=\frac 1 \beta \int d\epsilon  \, \frac {\partial p(\epsilon)}{\partial \epsilon} \left( \ln \sum_m e^{-\beta E_m(\mu-\epsilon)} \right)
\end{split}
\end{align} 
with $E_m(U,\mu')=U m(m-1)/2-\mu' m$.

In the specific case of box disorder with 
\begin{equation}
	\frac {\partial p(\epsilon)}{\partial \epsilon}=\frac{1}{\Delta}[\delta(\epsilon+\Delta/2)-\delta(\epsilon-\Delta/2)]
\end{equation}
the disorder averaged density takes on the form
\begin{equation}
\label{insulator_kappa}
	\overline n(\beta,U,\mu,\Delta)=\frac 1 {\Delta \beta} \ln \left[\frac{\sum_m e^{-\beta E_m(\mu+\Delta/2)}}{\sum_m e^{-\beta E_m(\mu-\Delta/2)}} \right]
\end{equation}
and the compressibility $\kappa=\frac{\partial \overline n}{\partial \mu}$ can be directly evaluated

\begin{align}\label{kappa_insulating_box}
\begin{split}
	\kappa=\frac{1}{\beta} \left(  \frac{\sum_m m\, e^{-\beta E_m(\mu+\Delta/2)} }{\sum_m e^{-\beta E_m(\mu+\Delta/2)}}  - \frac{\sum_m m\, e^{-\beta E_m(\mu-\Delta/2)} }{\sum_m e^{-\beta E_m(\mu-\Delta/2)}} \right).
\end{split}
\end{align} 

For speckle disorder, on the other hand, one has
\begin{equation}
		\frac {\partial p(\epsilon)}{\partial \epsilon}=\frac {\delta(\epsilon)}{\Delta} - \frac {\Theta(\epsilon)}{\Delta^2} e^{-\epsilon/\Delta}
\end{equation}
and the compressibility takes on the form
\begin{align}\label{kappa_insulating_speckle}
\begin{split}
	\kappa&=\frac{1}{\Delta}  \left[  \frac{ \sum_m m\, e^{-\beta E_m(\mu)} }{\sum_m  e^{-\beta E_m(\mu)}} \right.\\
	&\left.- \frac 1 \Delta \int_0^\infty d\epsilon \,e^{-\frac \epsilon \Delta} \frac{ \sum_m m\, e^{-\beta E_m(\mu-\epsilon)} }{\sum_m  e^{-\beta E_m(\mu-\epsilon)}} \right].
\end{split}
\end{align}

\section{Local Green's functions and DOS}
\label{app_loc_DOS}
To obtain the local DOS in an insulating phase, we calculate the
single particle Green's functions
\begin{align}\label{do_av_insulating_density}
\begin{split}
	G^>(t)&=\ev{b(t)b^\dag(0)}\\
	G^<(t)&=\ev{b^\dag(0)b(t)}\\
	G(t)&=-i[\Theta(t)\,G^>(t)+\Theta(-t)\,G^<(t) ]
\end{split}
\end{align} 
for local Fock states at finite temperature, where $b(t)$ is the on-site
particle annihilation operator in the Heisenberg representation. The
Fourier transformed Green's function can be calculated and takes on
the form
\begin{align}\label{Greens_fct_explicit}
\begin{split}
	\tilde G(\omega)&=\int_{-\infty}^\infty dt\, e^{i\omega t} G(t)\\
	=&\lim_{\gamma \searrow 0} \frac 1 {Z(\mu')} \sum_{m=0}^\infty e^{-\beta(\frac{U}{2}m(m-1)-\mu'm)}  \\ & \times  \left[(m+1) \frac{\omega - Um +\mu' -i \gamma}{(\omega - Um +\mu')^2+\gamma^2} \right.
	\\ &\left. -m \frac{\omega - U(m-1) +\mu' -i \gamma}{(\omega - U(m-1) +\mu')^2+\gamma^2}  \right],
\end{split}
\end{align} 
where $Z(\mu)=\sum_{m=0}^\infty e^{-\beta(\frac{U}{2}m(m-1)-\mu m)}$ is the local partition function.
This is related to the single particle DOS by

\begin{align}\label{DOS_explicit}
\begin{split}
	\rho(\omega,\mu')&=-\frac 1 \pi \mbox{Im}(\tilde G(\omega))\\
	&=\frac 1 {Z(\mu')} \sum_{m=0}^\infty e^{-\beta(\frac{U}{2}m(m-1)-\mu'm)}\left[ (m+1) \right. \\
	&\left. \times \delta(\omega-Um+\mu') + m \delta(\omega-U(m-1)+\mu')  \right].
\end{split}
\end{align} 

Averaging over the on-site energy distribution for speckle disorder
leads to the final expression
\begin{align}\label{DOS_fct_speckle}
\begin{split}
	\overline \rho(\omega,\mu,\Delta,\beta)=& \int d\epsilon \, p(\epsilon) \, \rho(\omega,\mu-\epsilon)\\
	=&\frac{1}{\Delta}\sum_{m=0}^{\infty}\frac{(m+1)\, \Theta(\omega-Um+\mu)}{ Z(Um-\omega)}  \\&\times e^{-\frac{\omega-Um+\mu}{\Delta}-\beta(m\omega -\frac U 2 m (m+1))}[1+e^{-\beta \omega}],
\end{split}
\end{align} 
whereas for a box disorder distribution one obtains
\cite{krutitsky-06njp187}
\begin{align}\label{DOS_fct_box}
\begin{split}
	\overline \rho(\omega,\mu,\Delta,\beta)=\frac{1}{\Delta}\sum_{m=0}^{\infty}(m+1)\frac{\Theta(\frac \Delta 2 -|\omega-Um+\mu|)}{ Z(Um-\omega)}  \\ \times \left[ e^{-\beta E_m(Um-\omega)}+ e^{-\beta E_{m+1}(Um-\omega)}\right].
\end{split}
\end{align} 

\section{Method of incorporating thermal fluctuations explicitly into SMFT}
\label{sec:explicit_thermal_fluctuations}
Instead of performing the thermal average before constructing the
conditional probability distribution for $\psi$, the SMFT furthermore
also allows to explicitly facilitate the thermal fluctuations of
$\psi$ in the probability distribution. We have not yet performed the numerical calculation, but will outline the procedure.

\begin{equation}
P(\psi|\eta)= \frac{d}{d \psi} \int d\epsilon \, p(\epsilon) \, \Theta\left[\psi- \frac{\Tr{b\, e^{-\beta H(\eta,\epsilon)} }}{\Tr{ e^{-\beta H(\eta,\epsilon)} }}       \right],
\end{equation}
where $\Theta(x)$ is the Heaviside step function.

This can be formulated in the following way: for fixed external
parameter $\eta$ (hence \emph{conditional} probability distribution)
the on-site energy is randomly drawn from $p(\epsilon)$ and the
resulting single site Hamiltonian is diagonalized. For the different
eigenstates $\ket{i(\epsilon,\eta)}$ with eigenenergies $E_i$ the
respective expectation value $\bra{i(\epsilon,\eta)} b
\ket{i(\epsilon,\eta)}$ is calculated. Within the 
grand canonical ensemble, this expectation value has the probability
$\left[ \Tr{ e^{-\beta H(\eta,\epsilon)} } \right]^{-1} \, e^{-\beta
  E_i}$ of occurring. This probability is uncorrelated to the probability
of a certain on-site energy $\epsilon$ occurring. Expressing this
mathematically, we obtain the explicit formula for the conditional
probability distribution
\begin{equation}
\begin{split}
P(\psi|\eta)= &\frac{d}{d \psi} \int d\epsilon \, p(\epsilon) \, \left[  \Tr{ e^{-\beta H(\eta,\epsilon)} }  \right]^{-1} \\
&\times \sum_{i=1}^\infty e^{-\beta E_i(\eta,\epsilon)} \,  \Theta\left[\psi- \bra{ i(\epsilon,\eta)} b \ket {i(\epsilon,\eta)}      \right]
\end{split}
\end{equation}
Equivalently on a formal level, one may work with the cumulative conditional probability function
\begin{equation}
\begin{split}
F(\psi|\eta)= & \int d\epsilon \, p(\epsilon) \, \left[  \Tr{ e^{-\beta H(\eta,\epsilon)} }  \right]^{-1} \\
&\times \sum_{i=1}^\infty e^{-\beta E_i(\eta,\epsilon)} \,  \Theta\left[\psi-  g_i(\mu-\epsilon,\eta)     \right]
\end{split}
\end{equation}
with 
\begin{equation}
	g_i(\mu-\epsilon,\eta)=\bra{ i(\epsilon,\eta)} b \ket {i(\epsilon,\eta)}.
\end{equation}
To evaluate the expression
\begin{equation}
	q_i(\psi)=\int d\epsilon \, f_i(\epsilon) \, \Theta(\psi-g_i(\mu-\epsilon,\eta))
\end{equation}
we perform a variable substitution
\begin{equation}
	\epsilon \mapsto x_i(\epsilon)
\end{equation}
and subsequently
\begin{equation}
	dx_i=\underbrace{\frac{d x_i(\epsilon)}{ d \epsilon}}_{f_i(\epsilon)} d \epsilon.
\end{equation}
Using the central theorem of calculus one can explicitly construct
\begin{equation}
	x_i(\epsilon)=\int_{-\infty}^\epsilon d\epsilon' \, f_i(\epsilon').
\end{equation}
Since 
\begin{equation}
	f_i(\epsilon)=p(\epsilon) \, \frac{e^{-\beta E_i(\epsilon,\eta)}}{\Tr{e^{-\beta H(\epsilon,\eta)}}}
\end{equation}
is a non-negative function the monotonously increasing function 
$x_i(\epsilon) \leftrightarrow \epsilon_i(x)$
is invertible in the relevant range. The conditional cumulative
density function can then be explicitly calculated from
\begin{equation}
	F(\psi|\eta)=\sum_{i=1}^{\infty}\int dx \, \Theta(\psi-g_i(\mu-\epsilon_i(x)),\eta)
\end{equation}
As in the SMFT for $T=0$, the self-consistency condition reads
\begin{equation}
	P(\psi)=\frac{d}{d\psi} \int_0^\infty d\eta \, Q(\eta) \, F(\psi|\eta),
\end{equation}
where $Q(\eta)$ is the $Z-$fold convolved and rescaled function of
$P(\psi)$ with the random variable $\eta=J\sum_{i=1}^Z \psi_i$.
The conditional cumulative density $F(\psi|\eta)$ now also contains the thermal fluctuations explicitly in the distribution (i.e. they are not averaged over within the self-consistency loop), so the subsequent determination of $P(\psi)$ is identical to the previous cases.

\end{appendix}


\end{document}